\documentclass[%
 aip,
 amsmath,amssymb,
 reprint,%
]{revtex4-1}

\usepackage[utf8]{inputenc}
\usepackage[T1]{fontenc}
\usepackage{xcolor}
\usepackage[normalem]{ulem}
\newcommand{\stkout}[1]{\ifmmode\text{\sout{\ensuremath{#1}}}\else\sout{#1}\fi}
\usepackage{etoolbox}
\usepackage{amsmath}
\usepackage{siunitx}
\usepackage{mathptmx}
\usepackage{multirow}
\usepackage{array}
\usepackage{makecell}
\usepackage{dcolumn}
\usepackage{graphicx}
\usepackage{xcolor}

\makeatletter
\def\@email#1#2{%
 \endgroup
 \patchcmd{\titleblock@produce}
  {\frontmatter@RRAPformat}
  {\frontmatter@RRAPformat{\produce@RRAP{*#1\href{mailto:#2}{#2}}}\frontmatter@RRAPformat}
  {}{}
}%
\makeatother
\begin{document}

\title{Interatomic and intermolecular Coulombic decay rates from 
equation-of-motion coupled-cluster theory with complex basis functions}
\author{Valentina Parravicini}
\author{Thomas-C. Jagau}
\affiliation{Department of Chemistry, KU Leuven, B-3001 Leuven, Belgium}


\begin{abstract} 
When a vacancy is created in an inner-valence orbital of a dimer of atoms 
or molecules, the resulting species can undergo interatomic/intermolecular 
Coulombic decay (ICD): the hole is filled through a relaxation process 
that leads to a doubly ionized cluster with two positively charged atoms 
or molecules. Since they are subject to electronic decay, inner-valence 
ionized states are not bound states but electronic resonances whose 
transient nature can only be described with special quantum-chemical 
methods. In this work, we explore the capacity of equation-of-motion 
coupled-cluster theory combined with two techniques from non-Hermitian 
quantum mechanics, complex basis functions and Feshbach-Fano projection, 
to describe ICD. To this end, we compute decay rates of several dimers: 
Ne$_2$, NeAr, NeMg, and (HF)$_2$, among which the energy of the outgoing 
electron varies between 0.3 eV and 16 eV. We observe that both methods 
deliver better results when the outgoing electron is fast, but the 
characteristic $R^{-6}$ distance dependence of the ICD width is captured 
much better with complex basis functions.
\end{abstract}


\maketitle


\section{Introduction} \label{section:intro}
\begin{figure*}[ht]
\centering
\makebox[\textwidth][c]{
\includegraphics[width=16cm, trim=20 0 0 0, clip]{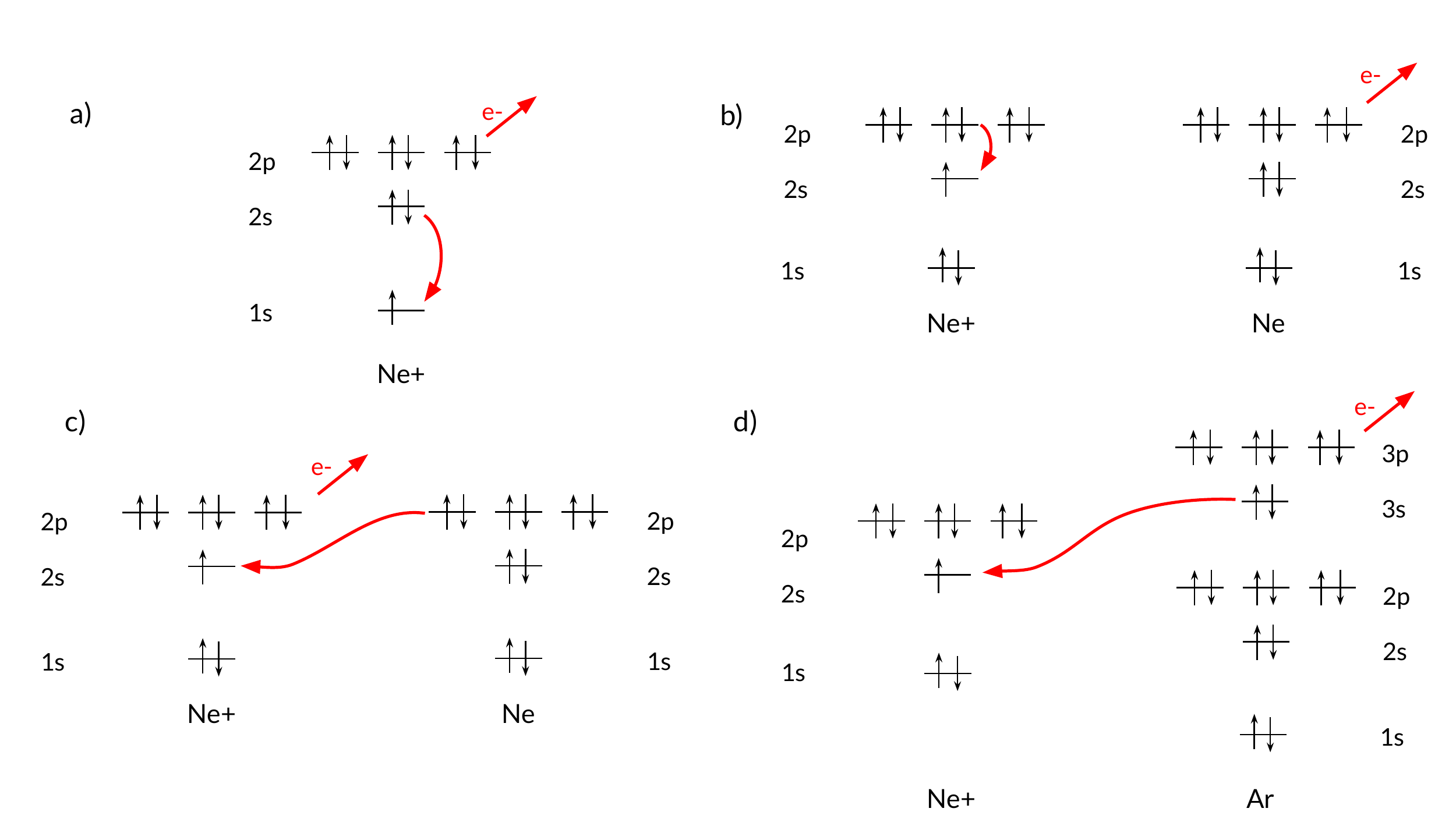}}
\caption{Schematic representation of non-radiative decay processes 
of core-ionized and inner-valence ionized species: a) Auger decay 
b) ICD – direct term c) ICD – exchange term d) ETMD}
\label{fig:decay_procs} \end{figure*}

Interatomic/intermolecular Coulombic decay (ICD)\cite{Cederbaum97,Jahnke20} 
is a very efficient non-radiative relaxation process that occurs in atomic 
and molecular clusters. When a vacancy has been created in an inner-valence 
orbital, non-radiative intramolecular relaxation, such as Auger-Meitner 
decay\cite{Meitner22,Auger23} (Fig. \ref{fig:decay_procs}a) cannot take 
place as it is energetically forbidden. The relaxation of the atom or 
molecule can, however, lead to the ionization of a neighbour. This 
mechanism is allowed when the lowest double ionization energy of the 
cluster is lower than the ionization energy that was needed to create 
the vacancy. This is the case for inner-valence orbitals, while 
outer-valence orbitals do not have enough energy to relax via this 
process. We note, however, that the distinction between inner-valence 
orbitals and outer-valence orbitals is not as obvious as between core 
and valence orbitals of light atoms. Rather, one has to compute the 
relevant energies explicitly to see if ICD is possible. 

ICD happens via two different mechanisms, referred to as direct and 
exchange.\cite{Santra01-2} The former (Fig. \ref{fig:decay_procs}b) 
consists in an energy transfer process, while the latter (Fig. 
\ref{fig:decay_procs}c) is mediated by the exchange of an electron 
between the two partners. For clusters of rare gas atoms, it has been 
argued that the direct contribution is more relevant because orbital 
overlap is small in those cases. The direct contribution is the 
one that is responsible for the characteristic distance dependence 
of the decay width of ICD, that is, $R^{-6}$ with $R$ being the 
interatomic distance. This result is obtained by expanding the 
relevant two-electron repulsion integral in an inverse power series 
in $R$.\cite{Thomas02,Averbukh04} The exchange contribution instead 
decays more rapidly with an exponential dependence on $R$ as the two 
partners are moved apart because it is driven by orbital overlap.

Typical ICD lifetimes are of the order of tens or hundreds of 
femtoseconds. ICD outperforms radiative decay pathways like photon 
emission, which usually takes place on the nanosecond scale, and 
is competitive with nuclear dynamics. Compared to Auger decay, 
however, ICD happens on a longer timescale; the former process 
typically has lifetimes of a few femtoseconds. For this reason, ICD 
widths are smaller than Auger decay widths. Also, ICD lifetimes have 
been proven to be strongly affected by environmental factors, for 
example, the presence of other neighbouring species\cite{Miteva17,
Votavova19}, their number and their structure.\cite{Ohrwall04,
Barth06,Ghosh14,Fasshauer16,Kumar22}. 

Similar to Auger decay, we can identify different doubly ionized 
states into which the initial inner-valence ionized state can 
decay. We will refer to them as target or final states. The energy 
differences between initial and final states are identical to the 
energy of the outgoing electron, which typically is in the range 
of 0--20 eV for ICD. This means that ICD electrons are at least 
one order of magnitude slower than Auger electrons, which usually 
have energies of hundreds of eV for first-row nuclei or even thousands 
of eV in the case of heavier nuclei. 

Closely related to ICD is electron transfer mediated decay\cite{Zobeley01} 
(ETMD), where the relaxation of an inner-valence vacancy results 
in a doubly ionized neighbour (Fig. \ref{fig:decay_procs}d). This 
process is allowed when the ionized atom has a neighbour with a 
double ionization energy lower than the ionization energy associated 
with the initial vacancy. ETMD is slower than ICD and only becomes 
relevant at very short internuclear distances. In fact, since ETMD 
is driven by orbital overlap, the decay width decreases with distance 
in the same way as the exchange term of ICD, that is, as $e^{-R}$.

Owing to their instability towards electron loss, inner-valence ionized 
states can be categorized as electronic resonances.\cite{Moiseyev11,
Jagau17,Jagau22} These states are not $L^2$-integrable and cannot 
be tackled with quantum-chemical methods suitable for bound states 
unless the decay is ignored. However, several methods have been 
devised to extend bound-state methods to capture the transient 
nature of resonances. In the framework of non-Hermitian quantum 
mechanics,\cite{Moiseyev11} the energy of resonances is expressed as 
\begin{equation}
E_\text{res} = E_R - i \, \Gamma/2,
\label{eq:energy} \end{equation}
where $E_R$ is the real part of the energy, also called the position 
of the resonance, and $\Gamma$ its decay width, related to the lifetime 
as $\tau$ = 1/$\Gamma$.

Several approaches have been introduced for the \textit{ab initio} 
computation of ICD and ETMD widths. In Feshbach-Fano resonance 
theory,\cite{Fano61,Feshbach62} the outgoing electron is modeled 
explicitly or, alternatively, Stieltjes imaging\cite{Langhoff03} 
is employed. The latter approach has been employed in conjunction 
with algebraic diagrammatic construction\cite{Averbukh05,Santra01-2,
Kolorenc20} (ADC) and configuration interaction\cite{Miteva17-2} 
(CI). Other options are R-matrix theory\cite{Sisourat17} and complex 
adsorbing potentials\cite{Santra01,Vaval07,Ghosh13} (CAPs), which 
decouple the resonance from the continuum and make it square-integrable. 

In this work, we explore the performance of two further non-Hermitian 
techniques for ICD widths. These are Feshbach-Fano resonance 
theory\cite{Fano61,Feshbach62} with a plane-wave description for 
the outgoing electron\cite{Skomorowski21} and the method of complex 
basis functions\cite{McCurdy78} (CBFs). We combine both techniques 
with equation-of-motion coupled-cluster singles and doubles (EOM-CCSD) 
and investigate ICD in the dimers Ne$_2$, NeAr, NeMg, and (HF)$_2$. 
EOM-CCSD\cite{Stanton93,Sneskov12} is a powerful \textit{ab initio} 
electronic-structure method to describe various types of excited 
states. In this work, we employ its ionization potential 
variant\cite{Stanton94} (EOM-IP-CCSD) to describe the initial 
states of ICD and its double ionization variant (EOM-DIP-CCSD) 
to describe the final states.\cite{Nooijen97,Sattelmeyer03} 
The CBF method has already been successfully employed to describe 
various other types of electronic resonances, such as temporary 
anions\cite{McCurdy78,McCurdy80,Rescigno80,Honigmann06,Honigmann10,
White15,White15-2,White17} and molecules in strong electric 
fields.\cite{Jagau18,Hernandez19,Hernandez20,Thompson19} Recently, 
CBFs combined with EOM-CCSD and CCSD were applied to molecular 
Auger decay.\cite{Matz22,Matz22-2,Jayadev23}  The combination 
of Feshbach-Fano theory with EOM-CCSD and an explicit representation 
of the outgoing electron has also been employed to compute molecular 
Auger decay widths.\cite{Skomorowski21,Skomorowski21-2,Jayadev23}

The remainder of the article is structured as follows: Sec. 
\ref{sec:th} summarizes the computational methods, while Secs. 
\ref{sec:ne2}--\ref{sec:hf2} present our numerical results for 
Ne$_2$, NeAr, NeMg, and (HF)$_2$, respectively. In Sec. 
\ref{sec:disc} we give our general conclusions.

\section{Theoretical and computational details} \label{sec:th}

\subsection{Method of complex basis functions} 
\label{section:theory_cbf}
The CBF method\cite{McCurdy78} is a variant of complex scaling\cite{
Moiseyev11,Aguilar71,Balslev71} that is compatible with the 
Born-Oppenheimer approximation, and thus suitable for molecules. 
The two are related through the equation
\begin{equation}
E_\text{res} = \frac{\langle \Psi(r) | \hat{H}(re^{i\theta}) | 
\Psi(r)\rangle}{\langle\Psi(r) | \Psi(r)\rangle} = \frac{\langle 
\Psi(re^{-i\theta}) | \hat{H}(r) | \Psi(re^{-i\theta}) \rangle}{\langle 
\Psi(re^{-i\theta}) | \Psi(re^{-i\theta}) \rangle}
\label{eq:cbf}
\end{equation} 
where the complex scaling angle $\theta$ can assume values between 
0 and $\pi/4$. In the CBF method, the complex-scaled wave function 
is expressed in terms of basis functions with scaled exponents, as 
in the context of Gaussian functions scaling the exponent is 
asymptotically equivalent to scaling the coordinates. Additional 
flexibility results from the fact that not all but only some 
functions in a basis set need to be scaled. A Hamiltonian built 
in such a basis set will have complex-valued eigenvalues that can 
be interpreted in terms of Eq. \eqref{eq:energy}.

Recently, it was shown that Auger decay widths can be treated with 
CBF versions of EOM-IP-CCSD and CCSD.\cite{Matz22,Matz22-2} The 
CCSD approach, which is based on a Hartree-Fock (HF) determinant 
for the decaying state is, however, not viable in the case of ICD. 
While it is, in general, no problem to construct a HF wave function 
with a hole in an inner-valence orbital using root-following 
techniques,\cite{Gilbert08} the CCSD wave function built from such 
a determinant suffers from variational collapse because the hole 
in the inner-valence shell is close in energy to occupied orbitals. 
This is different from core-ionized states, where the energetic 
separation between core and valence orbitals prevents a variational 
collapse. EOM-IP-CCSD does not suffer from such a problem because 
the HF and CCSD equations are solved for the neutral ground state. 

Unless specified otherwise, the basis set employed for all CBF 
calculations is aug-cc-pCVTZ(5sp): in this basis set, the triple 
$\zeta$ basis is modified by substituting the s and p shells with 
those from aug-cc-pCV5Z. It has been shown that this basis is 
suitable for the description of Auger decay.\cite{Matz22} When 
extra shells are added on top of aug-cc-pCVTZ(5sp), their exponents 
are obtained by recursively dividing the most diffuse exponent by 
a factor of 2, unless explicitly stated otherwise.

As concerns the exponents of the complex-scaled shells, it has been 
suggested that their value is related to the energy of the outgoing 
electron:\cite{Matz22} For the description of Auger electrons with a 
few hundreds of eV, one needs to scale shells with exponents in the 
range of 1--10, whereas diffuse shells are needed for temporary anions 
where the outgoing electron at most has a few eV. Consequently, for 
ICD, where electron energies are roughly in the same range as for 
temporary anions, we also expect diffuse complex-scaled shells to 
be important. 

This is investigated in detail for the Ne$_2$ dimer in Sec. 
\ref{sec:ne_cbf}. For other atoms, the exponents of the complex-scaled 
shells are adapted from neon according to the method from Ref. 
\citenum{Matz22}: a diffuseness factor $f$ is calculated for each 
atom, and the exponents ($\zeta$) for the new atom are calculated 
as $\zeta_2 = \zeta_1 \cdot f_2/f_1$. These factors are 6.36, 4.23, 
1.34, 0.41, 4.28 for the aug-cc-pCVTZ(5sp) basis sets of Ne, Ar, 
Mg, H and F, respectively.

A technical complication of CBF calculations is that one has to find 
an optimal scaling angle. In the present work, we determine the angle 
by minimizing $d(E_\text{res}-E_0)/d\theta$,\cite{Moiseyev78} where 
E$_0$ is the energy of the neutral ground state. This derivative is 
evaluated by recalculating the energy at different angles and numerical 
differentiation. The spacing between two angles is 2°, beside for the 
values reported in Table \ref{tab:ne_cbf}, where the spacing is 1.5°.

All CBF-EOM-IP-CCSD calculations presented in this work were carried 
out using a developer's version of the Q-Chem program package,\cite{qchem5} 
version 6.0. Details of the implementation are available from Refs. 
\citenum{White15,White15-2,White17,Zuev14}.




\subsection{Feshbach-Fano projection method} \label{section:theory_ff}

In Feshbach-Fano theory,\cite{Fano61,Feshbach62} resonances are 
described by the interaction of a bound state and scattering-like 
continuum states. The two sorts of states are obtained by 
partitioning the Hilbert space by means of projection operators, 
which can take many forms as long as they respect the condition 
to be mutually orthogonal and sum up to 1. 

Recently, Skomorowski and Krylov combined Feshbach-Fano theory 
with EOM-CCSD methods to describe Auger decay of core-ionized and 
core-excited states.\cite{Skomorowski21,Skomorowski21-2} In their 
approach, the bound part of the Hilbert space is defined using 
core-valence separation\cite{Cederbaum80,Coriani15,Vidal19} (CVS) 
and the continuum states are represented by products of EOM-CCSD 
states and a free-electron state, described by a continuum orbital.

In this work, we employ a similar approach to describe ICD. A 
generalized CVS scheme, where the ``core'' includes the inner-valence 
orbitals, is used to define the bound state. This state is decoupled 
from the continuum because the ICD final states are not included 
in the EOM-IP-CCSD excitation manifold. However, we note that the 
assumption underlying CVS, that core and valence orbitals can be 
treated separately, does not hold for inner-valence and outer-valence 
orbitals, as the energetic separation between them is usually just 
a few eVs or even less. The validity of this approach for treating 
ICD is thus questionable. 

We represent the continuum states as products of EOM-DIP-CCSD states 
and free-electron states described by plane waves. We note that 
it has been demonstrated in the framework of Auger decay that the 
plane-wave approximation is better when the energy of the outgoing 
electron is high.\cite{Skomorowski21} Since the energetic separation 
between initial and final states is much smaller for ICD than for 
Auger decay, the outgoing electron has much less energy, which again 
calls into question the validity of the approach.
 
All Feshbach-Fano-EOM-CCSD calculations presented in this work 
were carried out using a developer's version of the Q-Chem program 
package,\cite{qchem5} version 6.0. Details of the implementation 
are available from Ref. \citenum{Skomorowski21}. The basis set 
employed for all Feshbach-Fano-EOM-CCSD calculations is aug-cc-pCV5Z.






\section{Neon dimer} \label{sec:ne2}

\subsection{Energetics} \label{section:ne_en}
\begin{figure}[ht]
\includegraphics[width=9cm,trim=140 0 160 120,clip]{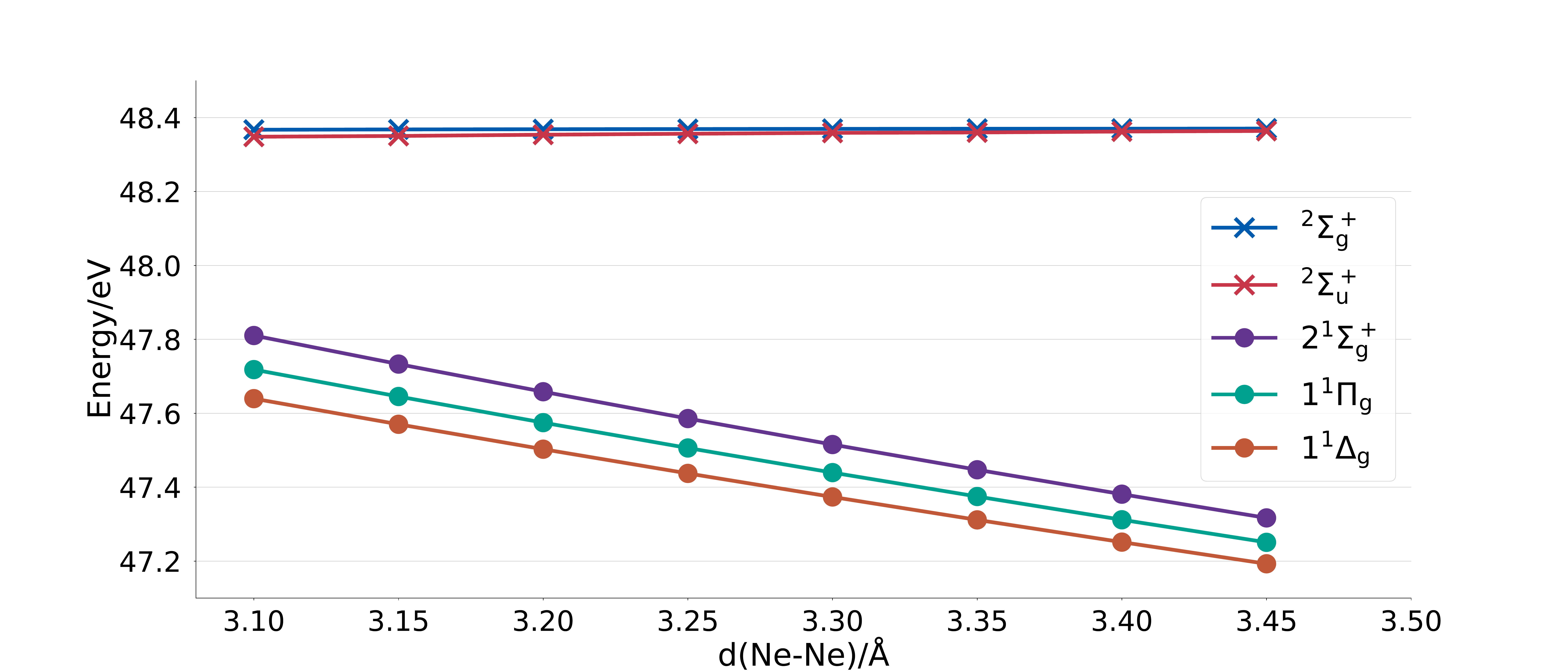}
\caption{Selected single and double ionization energies of 
Ne$_2$ as a function of the interatomic distance calculated 
with CBF-EOM-IP-CCSD/aug-cc-pCVTZ(5sp)+2s2p2d and 
EOM-DIP-CCSD/aug-cc-pCVTZ(5sp), respectively. Crosses correspond 
to resonances, thus initial states of ICD, and dots to final 
states of ICD.}
\label{fig:ne_ip} \end{figure}

The neon dimer has been investigated as a prototypical example 
of ICD in many works. Upon ionization of a 2s orbital, two states 
arise, $^2\Sigma_g^+$ and $^2\Sigma_u^+$, both of which can decay 
via ICD. At the equilibrium distance of neutral Ne$_2$ (3.2 \AA) 
the inner-valence ionization energies are 48.369 eV for the 
$^2\Sigma_g^+$ state and 48.354 eV for the $^2\Sigma_u^+$ state, 
meaning the two resonances are only 15 meV apart. In Fig. 
\ref{fig:ne_ip}, we report the ionization energies of the two 
states as a function of the interatomic distance. In the range 
considered (3.10--3.45 \AA), the change is less than 20 meV but 
we observe that the energy difference between the two states 
slightly decreases from 19 meV to 6 meV as the atoms are moved 
apart. 

We identified 12 possible ICD target states whose energies at 
the equilibrium distance are reported in Tab. \ref{tab:ne_dip}. 
We observe that all target states are close in energy spanning 
a range of less than 0.2 eV. Comparing the energy of initial 
and final states shows that the emitted ICD electron has an 
energy of less than 1 eV, which is in good agreement with 
previous theoretical\cite{Scheit04} and experimental\cite{Jahnke04} 
investigations that found the distribution of ICD electrons 
to peak at around 0.7 and 0.5 eV, respectively. 

We also studied the distance dependence of the double ionization 
energies of the target states. All 12 states remain accessible 
via ICD in the range considered here but, contrary to the initial 
states, the double ionization energies decrease by ca. 0.5 eV as 
the internuclear distance grows from 3.10 \AA\ to 3.45 \AA. This 
can be explained by the repulsion between the two positive 
charges being weaker when the two atoms are farther apart. 

The target states can be divided into three groups of almost 
degenerate states. In Fig. \ref{fig:ne_ip} we report one state 
for each group, while the results for all states can be found 
in the Supplementary Material. The lowest-lying group comprises 
6 states of $\Sigma$ and $\Delta$ symmetry from Tab. 
\ref{tab:ne_dip}, while the middle group comprises 4 $\Pi$ 
states and the highest-lying group 2 $\Sigma$ states.



\begin{table}[ht]
\caption{Selected single and double ionization energies of Ne$_2$ 
computed with EOM-DIP-CCSD/aug-cc-pCVTZ(5sp)+2s2p2d at $R=3.2$ \AA. 
Only doubly ionized states accessible via ICD are reported.}
\centering
\begin{ruledtabular}
{\begin{tabular}{ccc} 
Resonance state & Energy/eV  &  \\
\hline
$^2\Sigma_g^+$ & 48.369 & \\
$^2\Sigma_u^+$ & 48.354 & \\
\hline
Target state & Singlet/eV  & Triplet/eV \\
\hline
1 $\Delta_u$ & 47.502 & \\
1 $\Sigma_g^-$ &   & 47.502 \\
1 $\Delta_g$ & 47.503 & \\
1 $\Sigma_u^-$ &   & 47.503 \\
1 $\Sigma_g^+$ & 47.504 &  \\
1 $\Sigma_u^+$ &   & 47.504   \\
1 $\Pi_g$  & 47.575& 47.578 \\
1 $\Pi_u$  & 47.578 & 47.575 \\
2 $\Sigma_g^+$ & 47.658 & \\
2 $\Sigma_u^+$ &  & 47.658 \\
\end{tabular}}
\end{ruledtabular}
\label{tab:ne_dip}
\end{table}


\subsection{ICD widths computed with complex basis functions} 
\label{sec:ne_cbf}

In Tab. \ref{tab:ne_cbf} the total widths of the two states of 
Ne$_2^+$ calculated with CBF-EOM-IP-CCSD are reported. Several 
attempts were made, varying the size of the basis set and the number, 
angular momentum, and exponent of the complex-scaled functions. It 
can be observed that the widths of both states are very sensitive 
to the exponents of the complex-scaled shells. This especially applies 
to the \textit{ungerade} state whose decay width is most difficult 
to capture. 

Contrary to temporary anions, adding scaled functions that are more 
diffuse than the unscaled ones already present in the aug-cc-pCVTZ 
basis set does not lead to a better description of the resonance state. 
Decay widths are uniformly too small, in some cases even negative. 
The results improve, especially for the $^2\Sigma_g^+$ state, when 
we replace the s and p shells in the aug-cc-pCVTZ basis set with 
the ones from aug-cc-pCV5Z and scale the most diffuse shell of each 
angular momentum (s, p and d) in the resulting basis set. Adding new 
functions instead of scaling the ones already present in the basis set 
leads to further improvement. Notably, the exponents of the additional 
shells need not be smaller than the other exponents. This is somewhat 
similar to Auger decay even though the functions needed there have 
exponents in the range of 1--10.

As shown in Tab. \ref{tab:ne_lit}, the best match with other values 
computed at the same internuclear distance (3.2 \AA) is obtained with 
exponents 0.5 and 0.08 for the two extra s, p and d shells. We also 
note a Fano-CI study that reported widths of 5.0 and 9.0 meV for the 
\textit{gerade} and \textit{ungerade} states, respectively, at R=3.09 
\AA\ employing a cc-pVQZ+7s7p7d basis set.\cite{Miteva17-2} Also, the 
lifetime of the average of the two states was measured by means of a 
pump-probe experiment \cite{Schnorr13} as (150$\pm$50) fs, which 
corresponds to a width between 3.3 and 6.6 meV. It is, however, not 
possible to directly compare this latter value to computational 
results that do not take into account nuclear motion.



\begin{table*}[ht]
\caption{Total decay widths $\Gamma$ of the two Ne$^+_2$(2s$^{-1}$) 
states computed with CBF-EOM-IP-CCSD and various basis sets at $R=3.2$ 
\AA. Values in meV.}
\begin{ruledtabular}
\begin{tabular}{lcccc} 
Basis set &	Scaled basis functions & Exponents of the scaled 
functions & $^2\Sigma_g^+$ & $^2\Sigma_u^+$ \\ \hline
aug-cc-pCVTZ+2s2p & \makecell{extra shells + most diff. spd shell \\ 
of the predefined basis set} & \makecell{s (0.11, 0.57, 0.03) \\ 
p (0.09, 0.04, 0.03) d (0.39) } & 0.4 & 0.6 \\
aug-cc-pCVTZ+2s2p & extra shells & \makecell{s (0.57, 0.03) \\ 
p (0.04, 0.03)} & -2.2 & -0.4 \\
aug-cc-pCVTZ+2s2p2d & extra shells & \makecell{s (0.57, 0.03) p (0.04, \\ 
0.03) d (0.19, 0.10) } & 3.0 & 2.6 \\
aug-cc-pCVTZ(5sp) & \makecell{most diff. spd shell \\ 
of the predefined basis set} & \makecell{s (0.10) p (0.06) \\ 
d (0.39)} & 13.2 & 5.2 \\
aug-cc-pCVTZ(5sp) & \makecell{2 most diff. spd shell \\ 
of the predefined basis set} & \makecell{s (0.29, 0.10) p (0.19, \\ 
0.06) d (1.10, 0.39)} & 2.6 & 0.4 \\
aug-cc-pCVTZ(5sp) & \makecell{3 most diff. spd shell \\ 
of the predefined basis set} & \makecell{s (0.73, 0.29, 0.10) \\ 
p (0.52, 0.19, 0.06) \\ d (4.01, 1.10, 0.39)} & 6.6 & 2.4 \\
aug-cc-pCVTZ(5sp)+2s2p2d & extra shells & 1.0, 0.5 & 9.2 & 1.0 \\
aug-cc-pCVTZ(5sp)+2s2p2d & extra shells & 0.5, 0.2 & 11.6 & 1.6 \\
aug-cc-pCVTZ(5sp)+2s2p2d & extra shells & 0.2, 0.08 & 0.8 & 0.6 \\
aug-cc-pCVTZ(5sp)+2s2p2d & extra shells & 0.5, 0.08 & 10.1 & 7.6 \\
\end{tabular} \end{ruledtabular}
\label{tab:ne_cbf}
\end{table*}   

\begin{table}[ht]
\caption{Total decay width $\Gamma$ of the two Ne$^+_2$(2s$^{-1}$) 
states computed with various electronic structure methods at $R=3.2$ 
\AA. An experimental value is also given. All values are in meV.}
\begin{ruledtabular}
{\begin{tabular}{lcc} 
Method & $^2\Sigma_g^+$ & $^2\Sigma_u^+$ \\ \hline
CAP-CI/d-aug-cc-pVDZ +3s3p3d \cite{Santra01} &  & 10.3 \\
CAP-ADC(2)/d-aug-cc-pV5Z \cite{Vaval07} &  & 7.1 \\
CAP-EOM-CCSD/aug-cc-pCVTZ+3s3p3d \cite{Ghosh14}	&  & 7.2 \\
\makecell[l]{CBF-EOM-IP-CCSD/aug-cc-pCVTZ(5sp) \\ 
+2s2p2d (this work)} & 10.1 & 7.6 \\
\makecell[l]{Feshbach-Fano-EOM-IP-CCSD/ \\ 
aug-cc-pCV5Z (this work)} & 8.2 & 7.3 \\
\multicolumn{3}{l}{Experimental value \cite{Schnorr13} 
\qquad\qquad\quad (150 $\pm$ 50) fs $\widehat{=}$ 3.3--6.6 meV} \\
\end{tabular}} \end{ruledtabular}
\label{tab:ne_lit}
\end{table}


\begin{figure}[h] 
\includegraphics[width=7cm]{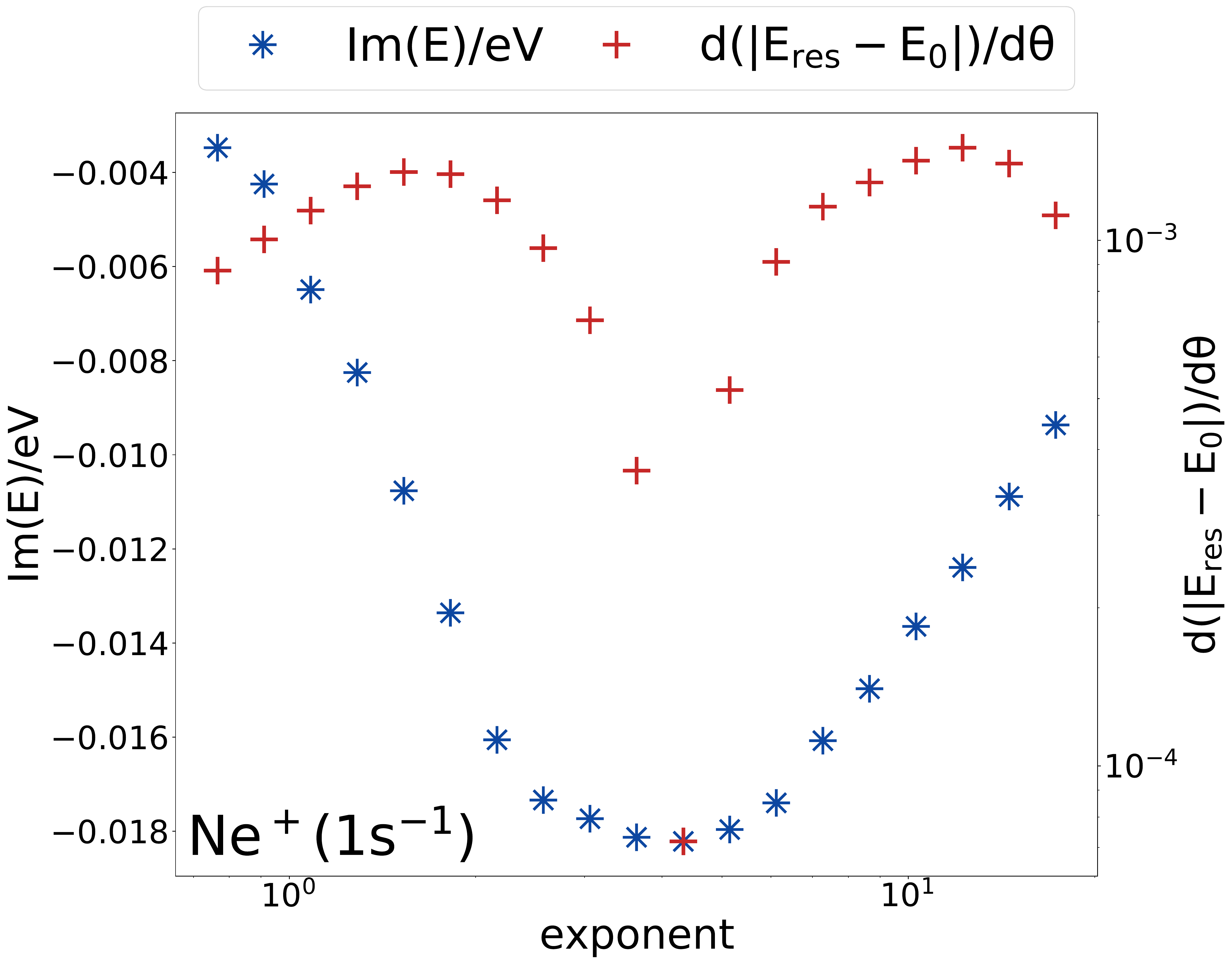} \\ [0.5cm] 
\includegraphics[width=7cm]{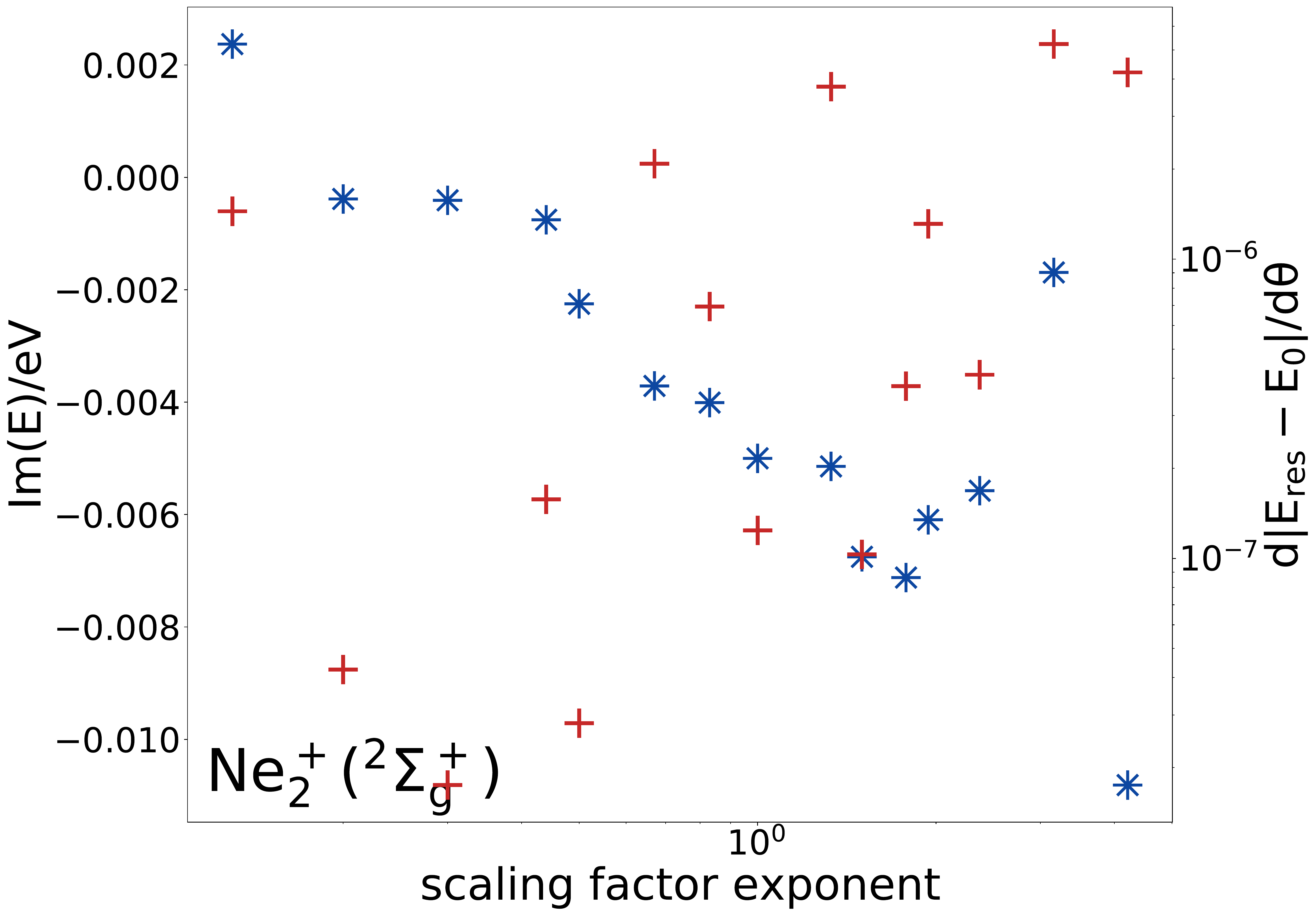} \\ [0.5cm]
\includegraphics[width=7cm]{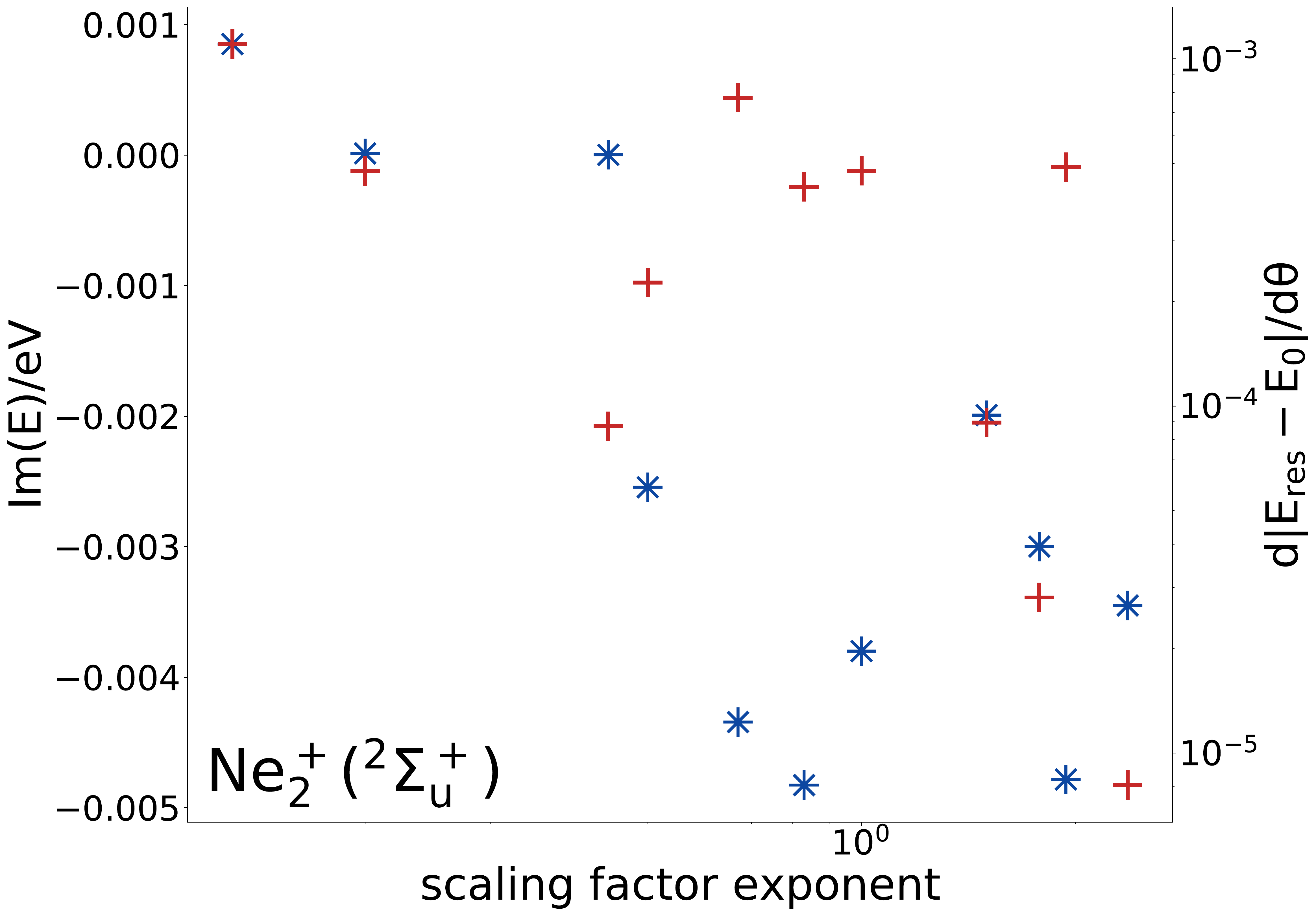} 
\caption{Imaginary energy ($=-\Gamma/2)$ and $d(E_\text{res}-E_0)/d\theta$ 
for the resonance states Ne$^+$(1s$^{-1}$) \cite{Matz22}, Ne($^2\Sigma_g^+$) 
(this work) and Ne($^2\Sigma_u^+$) (this work) as a function of the 
value of the exponents of the complex-scaled functions.}
\label{fig:ne_exp}
\end{figure}

Starting from the best result in Tab. \ref{tab:ne_cbf}, i.e., the values 
in the last row, we explored the sensitivity of the ICD widths to small 
changes in the exponents. Such an analysis had been performed for Auger 
decay\cite{Matz22} and it had been found that the optimal exponents 
are marked by concurrent minima in $\text{Im}(E_{res}-E_0$) and 
$d(E_{res}-E_0)/d\theta$. This is reproduced in the top panel of Fig. 
\ref{fig:ne_exp}. In the middle and lower panels of Fig. \ref{fig:ne_exp} 
we report the results of the corresponding analysis for the $^2\Sigma_g^+$ 
and $^2\Sigma_u^+$ states of Ne$_2^+$. We varied the exponents of s, p, 
and d shells together; their values for each point are reported in the 
Supplementary Material. For each new basis set, the optimal complex-scaling 
angle was determined anew. 

The stark difference between Auger decay and ICD is obvious in Fig. 
\ref{fig:ne_exp}. While there are clear concurrent minima in both 
curves for Auger decay, the plots are a lot less neat for ICD. For 
the \textit{gerade} state, there is a minimum in Im($E$) but this is 
not the case for the \textit{ungerade} state. Moreover, the quantity 
$d(E_{res}-E_0)/d\theta$, which is used to find the optimal complex 
scaling angle, is not a good indicator of the quality of the exponents 
in the case of ICD. A possible explanation may be that the ICD widths 
of Ne$_2^+$ are at least one order of magnitude smaller than the Auger 
decay widths studied in Ref. \onlinecite{Matz22}. 

The exponents that correspond to a minimum in $\text{Im}(E_\text{res} 
- E_0)$ for the $^2\Sigma_g^+$ state are 0.889 and 0.143 and yield decay 
widths of 14.2 meV for the $^2\Sigma_g^+$ state and 6.0 meV for the 
$^2\Sigma_u^+$ state. This is markedly different from the widths of 
10.1 and 7.6 meV obtained with the original exponents (0.5 and 0.08). 
The latter set of values is in better agreement with the other values 
compiled in Tab. \ref{tab:ne_lit} but a final decision which set of 
values is superior remains difficult. Notably, the two resonance states 
do not overlap with either set of $\Gamma$ values as the energy gap 
between them (15 meV) is bigger than the sum of their half widths. 

\begin{figure}[ht]
\includegraphics[width=9cm]{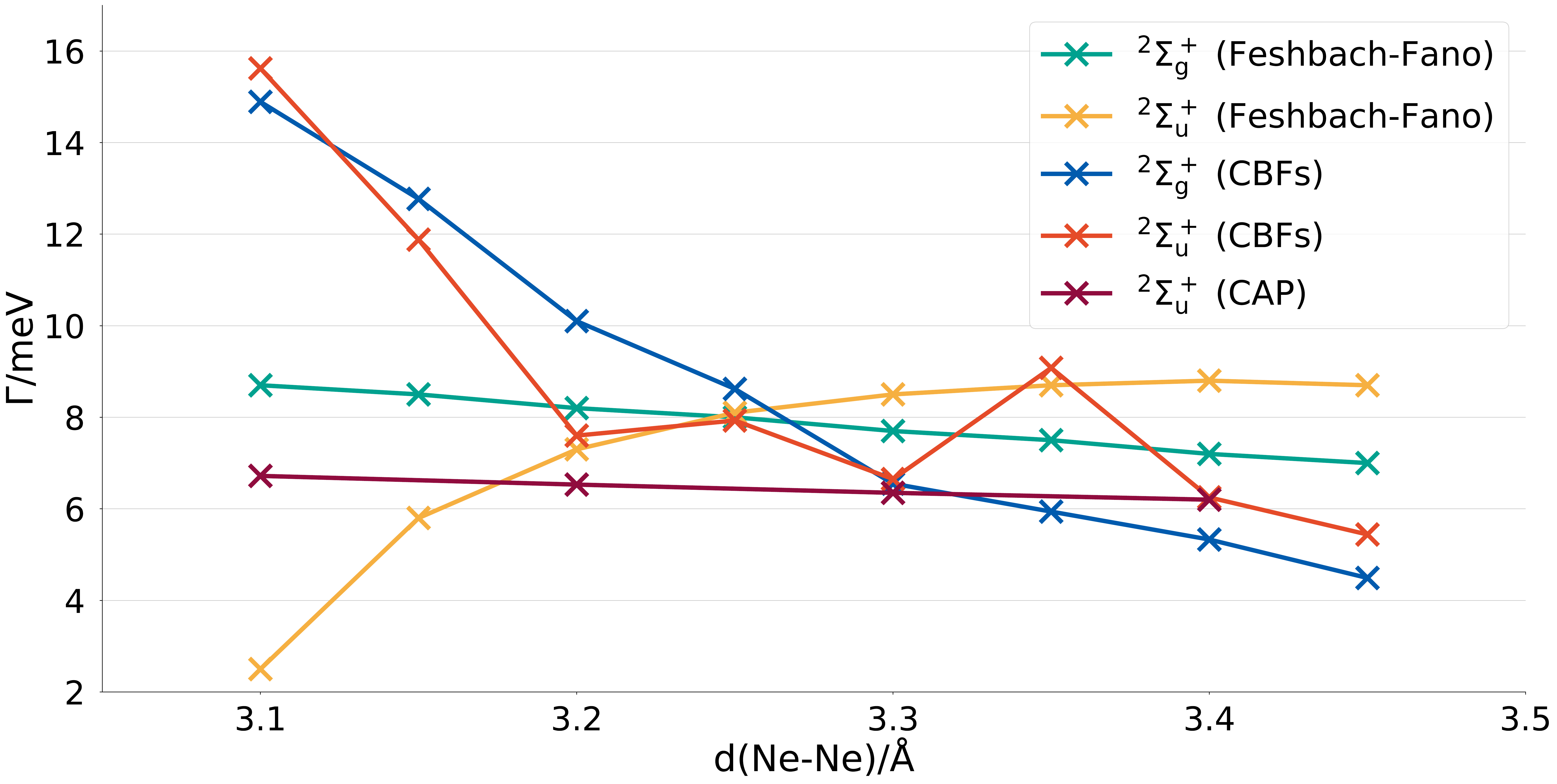}
\caption{Total decay width $\Gamma$ of the two Ne$^+_2$(2s$^{-1}$) 
states as a function of the interatomic distance calculated with 
CBF-EOM-IP-CCSD/aug-cc-pCVTZ(5sp)+2s2p2d (this work), 
Feshbach-Fano-EOM-IP-CCSD/aug-cc-pCV5Z (this work) and 
CAP-EOM-CCSD/aug-cc-pCVTZ (Ref. \onlinecite{Ghosh14}).}
\label{fig:ne_w} \end{figure}

We also investigated the distance dependence of the width of both states, 
this was done using the exponents from the last row of Tab. \ref{tab:ne_cbf}. 
The results are reported in Fig. \ref{fig:ne_w}. We see that the $R^{-6}$ 
dependency is not captured correctly; rather, a log-log plot reveals that 
the width of the $^2\Sigma_g^+$ state decreases as $R^{-12}$ and that of 
the $^2\Sigma_u^+$ as R$^{-8}$. Also, Fig. \ref{fig:ne_w} illustrates 
again that the \textit{ungerade} state is more difficult to describe 
than the \textit{gerade} state. We note that Ghosh \textit{et al.} also 
studied the distance dependence of width of the $\Sigma_u^+$ state using 
CAP-EOM-CCSD.\cite{Ghosh14} Our results are in good agreement at $R=3.2$ 
\AA\ and larger distances, but when the two atoms are closer to each other, 
we obtain larger widths than the ones reported in Ref. \onlinecite{Ghosh14}.

\subsection{ICD widths computed with the Feshbach-Fano method} 
\label{section:ne_ff}

As concerns Feshbach-Fano EOM-CCSD, we present ICD widths for Ne$_2^+$ 
computed at the equilibrium distance (3.2 \AA) with various basis 
sets in Tab. \ref{tab:ne_ff}. The width narrows as the basis set 
increases in size, and the value computed with aug-cc-pCV5Z compares 
well with the one obtained with CBF-EOM-IP-CCSD and previous 
computational results as one can see from Tab. \ref{tab:ne_lit}. 
However, as evident from Fig. \ref{fig:ne_w}, the method is not able 
to describe the distance dependence of the width correctly: instead 
of a decrease proportional to $R^{-6}$, we observe a linear trend 
for the $^2\Sigma_g^+$ state, whereas the width of the $^2\Sigma_u^+$ 
state even increases with the distance. 

We conclude that the Feshbach-Fano EOM-CCSD method devised for Auger 
decay is not suitable for ICD. This failure is not surprising given 
the shortcomings discussed in Sec. \ref{section:theory_ff}: first, our 
$Q$ projector for the bound part of the resonance is based on a 
generalization of CVS that does not hold for inner-valence orbitals 
as their energies are not well separated from the outer-valence 
orbitals. Moreover, we model the outgoing electron as a plane wave, 
which is an approximation that holds better when the energy of the 
outgoing electron is higher. 

\begin{table}[ht]
\caption{Total decay width of the Ne$^+_2$(2s$^{-1}$) states 
calculated with Feshbach-Fano-EOM-CCSD with various basis sets. 
Values are in meV.}
\begin{ruledtabular}
{\begin{tabular}{lcc} 
Basis set &	$^2\Sigma_g^+$ &	$^2\Sigma_u^+$ \\
\hline
aug-cc-pCVDZ & 11.2 & 19.2 \\
aug-cc-pCVTZ & 9.2 & 12.3 \\
aug-cc-pCVQZ & 8.5 & 8.8 \\
aug-cc-pCV5Z & 8.2 & 7.3 \\
aug-cc-pCVTZ(5sp) & 9.1 & 11.3 \\
\end{tabular}}
\end{ruledtabular}
\label{tab:ne_ff}
\end{table}



\section{Neon Argon dimer} \label{sec:near}

\subsection{Energetics}
\begin{figure}[ht]
\includegraphics[width=9cm]{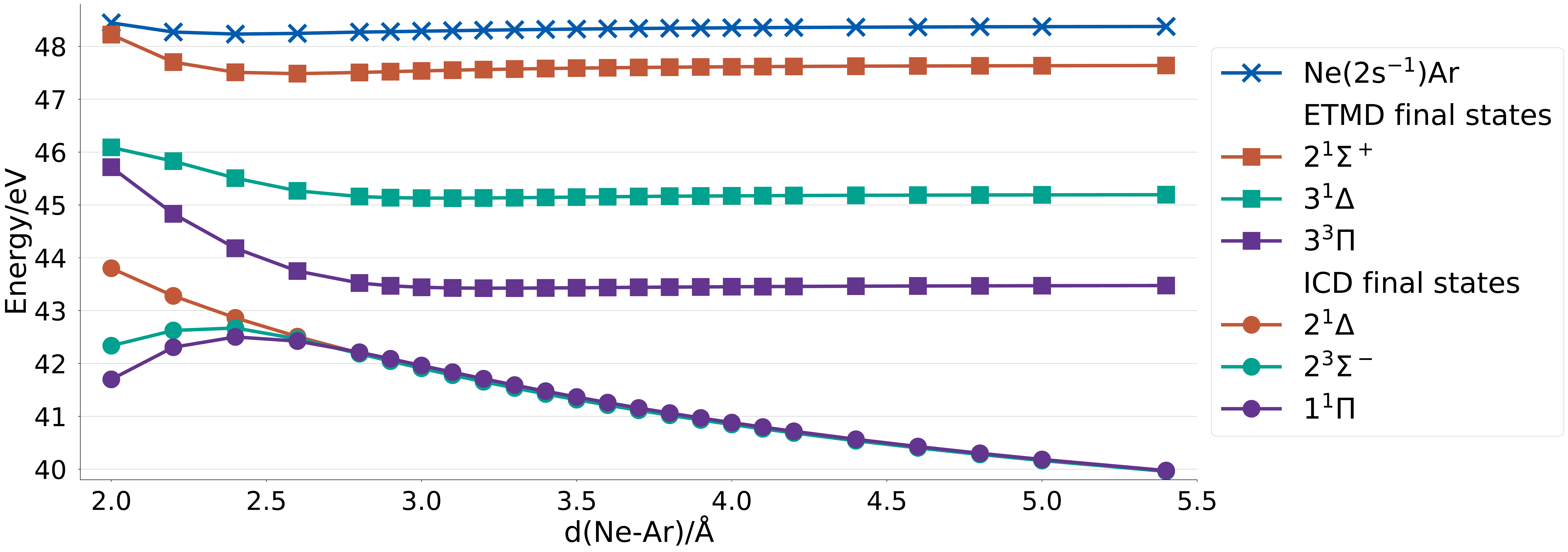}
\caption{Selected ionization and double ionization energies of 
NeAr as a function of the interatomic distance computed with 
CBF-EOM-IP-CCSD/aug-cc-pCVTZ(5sp)+2s2p2d and 
EOM-DIP-CCSD/aug-cc-pCVTZ(5sp), respectively. Crosses correspond 
to the resonance, squares and dots to ETMD and ICD final states, 
respectively.}
\label{fig:near_ip}
\end{figure}

As a second example, we study the NeAr dimer. It has been 
demonstrated that NeAr can undergo both ICD and ETMD upon 
ionization of the 2s orbital of the neon atom.\cite{Zobeley01} 
The two processes differ in the atoms that carry the vacancies 
in the final doubly ionized state: ICD results in Ne$^+$Ar$^+$ 
while in ETMD the two vacancies are found on Argon, NeAr$^{2+}$. 
Target states of the type Ne$^{2+}$Ar are higher in energy than 
the initial state and thus not accessible. In Fig. \ref{fig:near_ip}, 
the inner-valence ionization energy is plotted as a function of 
the interatomic distance (2.0--5.4 \AA). It has a very shallow 
minimum at 2.4 \AA\ and is otherwise almost independent of the 
interatomic distance, with a value of ca. 48.3 eV. Notably, the 
equilibrium distances of the neutral and ionized dimers are much 
larger, 3.5 \AA\ and 3.7 \AA, respectively. 

We identified 10 ICD target states and 6 ETMD target states. 
Their EOM-DIP-CCSD energies are reported in Tab. \ref{tab:near_dip}. 
Interestingly, all ICD final states are lower in energy than the 
ETMD target states. This is reflected in the energy of the ejected 
electron: in the case of ICD, it is around 7 eV for all target 
states, whereas in the case of ETMD it can assume values between 
0.6 eV and 5 eV depending on the target state. Notably, singlet 
and triplet ICD target states are almost degenerate, while this 
is not the case for ETMD target states. It is also worth mentioning 
that the energy of the ICD electrons is one order of magnitude 
larger for NeAr than for Ne$_2$. This is because the target states 
are more stabilized in the NeAr dimer, owing to the larger energy 
differences between the orbitals of Ne and Ar. The initial states, 
on the contrary, have almost the same ionization energies in Ne$_2$ 
and NeAr. Our results confirm the ICD electron energy distribution 
reported in Ref. \onlinecite{Scheit06}, where the kinetic energy 
spectra computed for singlet and triplet states both have a peak 
in intensity close to 7 eV. 


In Fig. \ref{fig:near_ip}, selected double ionization energies 
are reported as a function of the interatomic distance. Results 
for all states can be found in the Supplementary Material. In 
the range of distances considered (2.0--5.4 \AA) all target 
states remain accessible via both decay processes, although their 
relative energy varies. Similar to Ne$_2$, the double ionization 
energies of the ICD target states decrease as the two atoms are 
moved apart. However, the ICD target states of NeAr all converge 
to the same energy at large distances, which is not the case for 
Ne$_2$ where the states remain parallel to each other. 

The energies of the ETMD target states converge to three different 
values at large distances; in Fig. \ref{fig:near_ip}, one state 
from each group is reported. All double ionization energies 
corresponding to ETMD target states exhibit a shallow minimum at 
distances below the equilibrium distance of the ground state (3.5 
\AA). Since the two charges are on the same atom in the ETMD target 
states, the interatomic distance does not have a big impact on the 
double ionization energy, contrary to ICD target states. We note 
that our results are in agreement with curves that were computed 
with ADC(2).\cite{Zobeley01}


\begin{table}[ht]
\caption{Selected ionization and double ionization energies of 
NeAr computed with EOM-CCSD/aug-cc-pCVTZ(5sp)+2s2p2d.}
\centering
\begin{ruledtabular}
{\begin{tabular}{ccc} 
Resonance state & Energy / eV  &  \\
\hline
Ne(2s$^{-1}$)Ar$^+$ & 48.328 & \\
\hline
ICD final states  & Singlet / eV  & Triplet / eV \\     
\hline
1 $\Delta$ & 41.315 &    \\
1 $\Sigma^-$ &  &   41.315 \\
2 $\Delta$ & 41.317 &    \\
2 $\Sigma^-$ &  &   41.317  \\
1 $\Pi$ & 41.365 &   41.362 \\
2 $\Pi$ & 41.469 &   41.469 \\
1 $\Sigma^+$ & 41.529 &   41.534 \\
\hline
ETMD final states  & Singlet / eV  & Triplet / eV \\     
\hline
2 $\Sigma^+$ & 45.147 &    \\
3 $\Pi$  & 45.147 & 43.433    \\
3 $\Delta$ & 45.148 &   \\
3 $\Sigma^-$ & & 43.430 \\
3 $\Sigma^+$ & 47.589 &   \\
\end{tabular}}
\end{ruledtabular}
\label{tab:near_dip}
\end{table}


\subsection{Decay widths}
The decay width of Ne$^+$Ar has been computed several times: with 
Fano-ADC, a value of 17 meV was determined at the equilibrium distance 
of neutral NeAr (3.5 \AA)\cite{Scheit06}. A previous Fano-ADC treatment 
resulted in a value of 22 meV,\cite{Zobeley01} whereas CAP-EOM-CCSD 
computations delivered a value of 37.4 meV.\cite{Ghosh14} Also, it 
has been argued that the ETMD contribution to the width is several 
orders of magnitude smaller and decays more rapidly with the 
interatomic distance than the ICD contribution.\cite{Scheit06} 

\begin{figure}[ht]
\includegraphics[width=9cm]{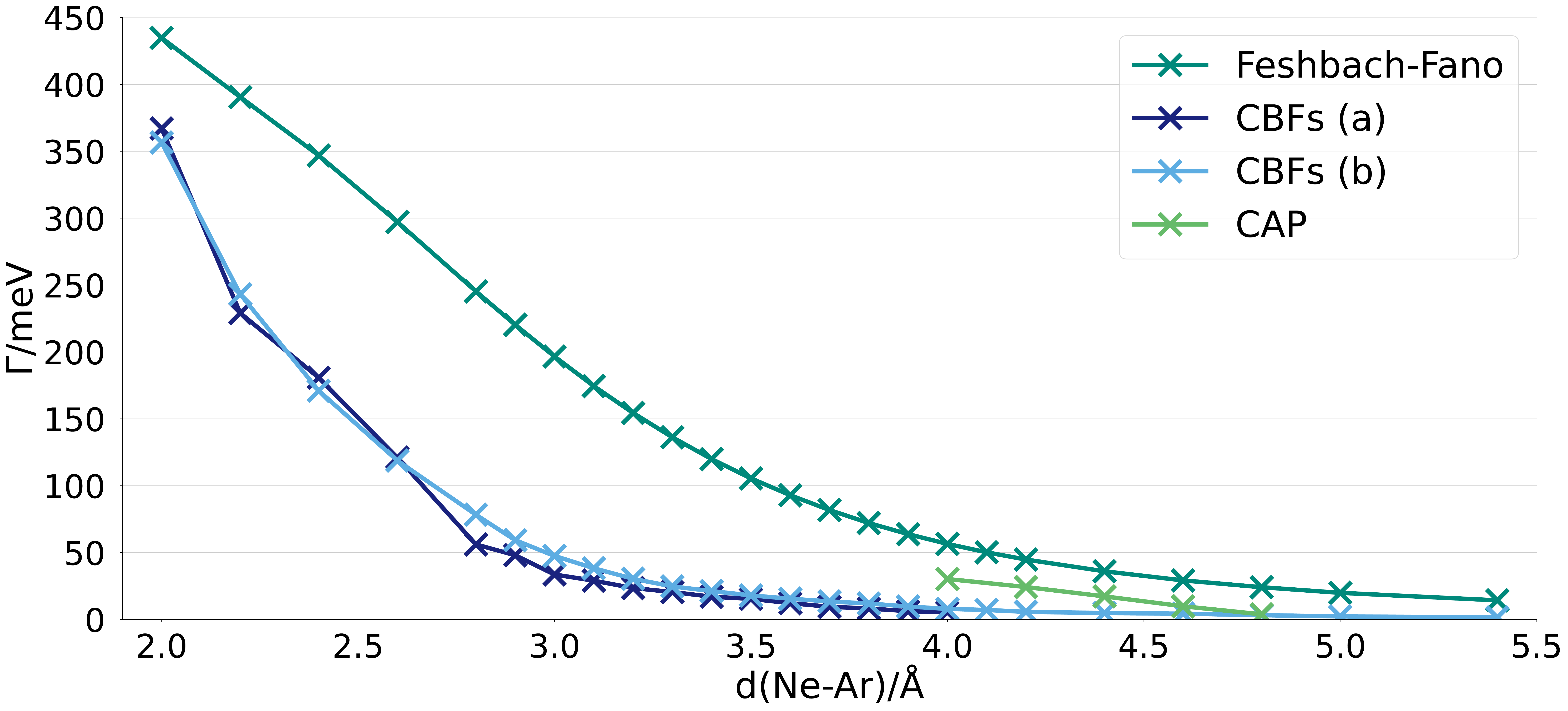}
\caption{Total decay width of Ne(2s$^{-1}$)Ar$^+$ as a function 
of the interatomic distance computed with 
CBF-EOM-CCSD/aug-cc-pCVTZ(5sp)+2s2p2d (this work), 
Feshbach-Fano-EOM-IP-CCSD/aug-cc-pCV5Z (this work) and 
CAP-EOM-CCSD/aug-cc-pCVTZ (Ref. \onlinecite{Ghosh14}). The labels 
(a) and (b) refer to results obtained with different values for the 
exponents of the complex-scaled shells of the argon atom.}
\label{fig:near_w}
\end{figure}

The total widths obtained in our CBF-EOM-IP-CCSD calculations 
comprises ICD and ETMD contributions. In Fig. \ref{fig:near_w}, 
we report the values computed at interatomic distances between 
2.0 \AA\ and 5.4 \AA. We employed in all calculations the 
aug-cc-pCVTZ(5sp)+2s2p2d basis set, which yielded the best 
results for Ne$_2^+$. For the neon atom, the exponents of the 
extra shells are those resulting from the optimization of the 
$^2\Sigma_g^+$ state of Ne$_2^+$, that is, 0.889 and 0.143, 
while for argon, two different sets of exponents were employed 
to check the sensitivity of the method: the results labeled (a) 
in Fig. \ref{fig:near_w} were computed using exponents that were 
obtained from the ones for Ne$_2^+$ using the procedure described 
in Sec. \ref{section:theory_cbf} resulting in values of 0.593 and 
0.095. The results labeled (b) in Fig. \ref{fig:near_w} were 
obtained with exponents of 0.944 and 0.059 for argon. Since the 
energy of the outgoing electron does not vary significantly in 
the range of distances considered here, we can use the same 
exponents for all internuclear distances. 

We find the total decay width of Ne(2s$^{-1}$)Ar$^+$ at 3.5 \AA\ 
to be 15.4 meV with basis set (a) and 18.0 meV with basis set (b), 
which shows that the width of Ne(2s$^{-1}$)Ar$^+$ is much less 
sensitive to the exponents of the basis functions than that of 
Ne$_2^+$. Notably, our results are in good agreement with Fano-ADC 
but considerably smaller than the CAP-EOM-CCSD value.

Given that the ETMD width is so much smaller than the ICD width, 
we do not expect it to have a significant effect on the distance 
dependence of the total width, meaning we expect $\Gamma(R) \sim 
R^{-6}$. Our CBF-EOM-IP-CCSD calculations yield $R^{-6.29}$ with 
basis set (a) and $R^{-5.72}$ with basis set (b). With basis set 
(a), $\Gamma$ becomes so small (5.4 meV) beyond 4.0 \AA\ that no 
minimum can be observed in $d(E_\text{res}-E_0)/d\theta$. We therefore 
re-examined the distance dependence removing the last 5 points, which 
improved the result to $\Gamma(R) \sim R^{-6.08}$. This might be due 
to the fact that CBF methods struggle to capture very small widths. 
Also with basis set (b), $\Gamma(R) \sim R^{-5.99}$ is obtained when 
the two widths at the shortest distance and the three at the largest 
distance are not taken into account. With this basis set, the maximum 
bond length at which a minimum in $d(E_\text{res}-E_0)/d\theta$ can 
be obtained is 5.4 \AA, after which the width again becomes too small 
to be described.

Contrary to the good performance of CBF-EOM-IP-CCSD, the 
Feshbach-Fano-EOM-CCSD method does not capture the distance dependence 
of the width of Ne(2s$^{-1}$)Ar$^+$ correctly. As Fig. \ref{fig:near_w} 
illustrates, $\Gamma$ is severely overestimated as compared to 
CBF-EOM-IP-CCSD. Although the values decrease when the atoms are moved 
apart, the further analysis yields $\Gamma(R)\sim R^{-3.72}$. This can 
be slightly improved to $R^{-4.65}$ by only considering values larger 
than 3.4 \AA, but this is still not a satisfactory result.

\section{Neon Magnesium dimer} \label{sec:nemg}
\subsection{Energetics}

\begin{figure}[ht]
\includegraphics[width=9cm]{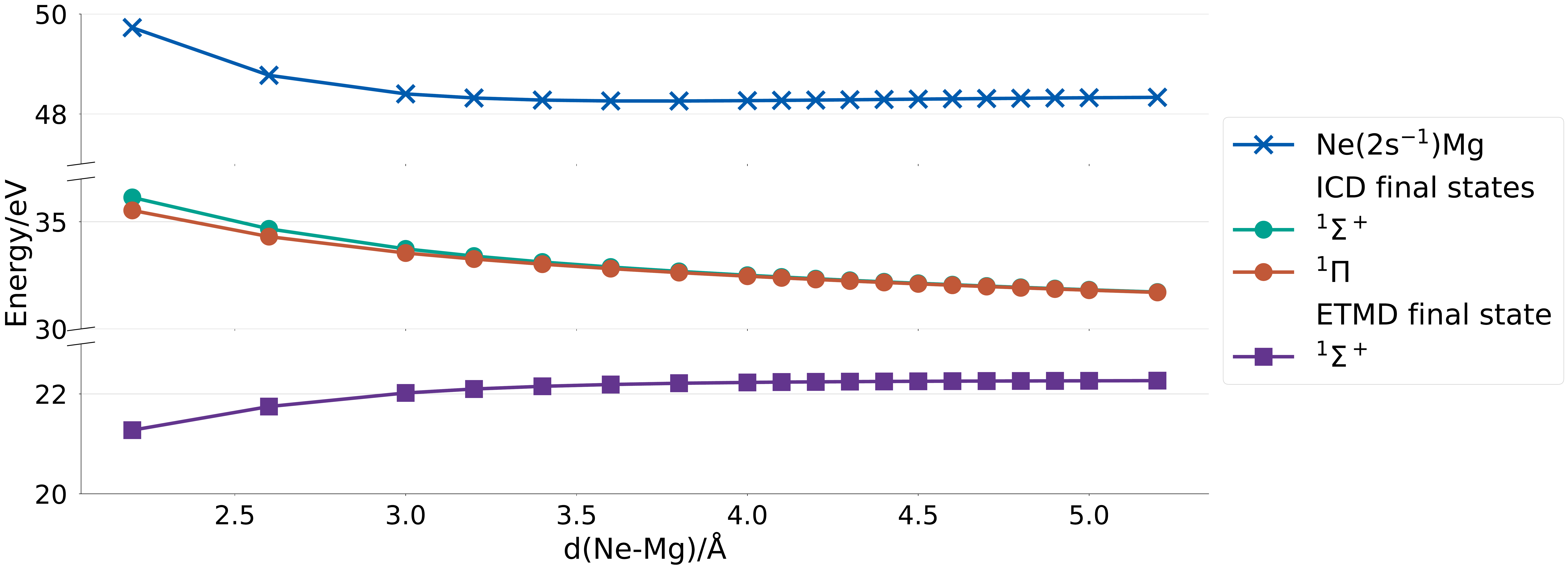}
\caption{Selected ionization and double ionization energies of 
NeMg as a function of the interatomic distance computed with 
CBF-EOM-IP-CCSD/aug-cc-pCVTZ(5sp)+2s2p2d and 
EOM-DIP-CCSD/aug-cc-pCVTZ(5sp), respectively. Crosses correspond 
to the resonance, squares and dots to ETMD and ICD target states, 
respectively.}
\label{fig:nemg_ip}
\end{figure}

As a third example of an atomic dimer, we consider NeMg. As in 
the case of NeAr, the relaxation of the 2s$^{-1}$ vacancy in the 
neon atom of NeMg can occur both via ICD and ETMD, with the latter 
process having a width of orders of magnitude smaller than the 
one of ICD.\cite{Averbukh04} The inner-valence ionization energy 
is ca. 48.3 eV, which is very similar to the previous examples. 
In Fig. \ref{fig:nemg_ip}, it is plotted as a function of the 
interatomic distance. The ionization energy exhibits a very 
shallow minimum at around 3.8 \AA, whereas the equilibrium 
distance of the neutral dimer is 4.4 \AA. 



For this system, we identified 4 target states accessible via ICD 
and 1 accessible via ETMD. In Tab. \ref{tab:nemg_dip}, we report 
their double ionization energies. ICD target states of the type 
Ne$^+$Mg$^+$ are ca. 16 eV lower in energy than the resonance state, 
which means that the ICD electron has twice as much energy as in 
the case of NeAr. Singlet and triplet final states are again almost 
degenerate. 

In the case of ETMD, which results in NeMg$^{++}$, the energy of the 
outgoing electron is much larger, close to 26 eV. It is interesting 
to note that, contrary to NeAr, the ETMD final state of NeMg is lower 
in energy than the ICD final states. This is because the double 
ionization of the Mg 3s orbital leads to a very stable electronic 
configuration, which outweighs that the two charges are localized 
on the same atom. Similar to NeAr, target states of the type 
Ne$^{++}$Mg are not accessible because they are too high in energy. 

All target states remain accessible at all distances considered here 
(2.2--5.2 \AA). As one can see from Fig. \ref{fig:nemg_ip}, where we 
report the double ionization energies of the singlet states as a 
function of the distance, the ICD final states decrease in energy 
as the two atoms are moved apart. This is consistent with the trend 
we observed for Ne$_2$ and NeAr. The double ionization energy of 
the ETMD state instead increases with distance, contrary to what 
we observed for the ETMD final states of NeAr. Also, the double 
ionization energy of the ETMD state does not exhibit a minimum for 
the distances considered here. Results for the ICD triplet states 
can be found in the Supplementary Material.

\begin{table}[h]
\caption{Selected ionization and double ionization energies of 
NeMg computed with EOM-CCSD/aug-cc-pCVTZ(5sp)+2s2p2d.}
\centering
\begin{ruledtabular}
{\begin{tabular}{ccc} 
Resonance state & Energy/eV  &  \\
\hline
Ne(2s$^{-1}$)Mg$^+$ & 48.291 & \\
\hline
ICD final state & Singlet/eV & Triplet/eV \\     
\hline
2 $\Sigma^+$ & 32.198 &   32.192 \\
1 $\Pi$ & 32.169 &   32.168 \\
\hline
ETMD final state & Singlet/eV & Triplet/eV \\     
\hline
1 $\Sigma^+$ & 22.249 & \\
\end{tabular}}
\end{ruledtabular}
\label{tab:nemg_dip}
\end{table}


\subsection{Decay widths}

\begin{figure}[ht]
\includegraphics[width=9cm]{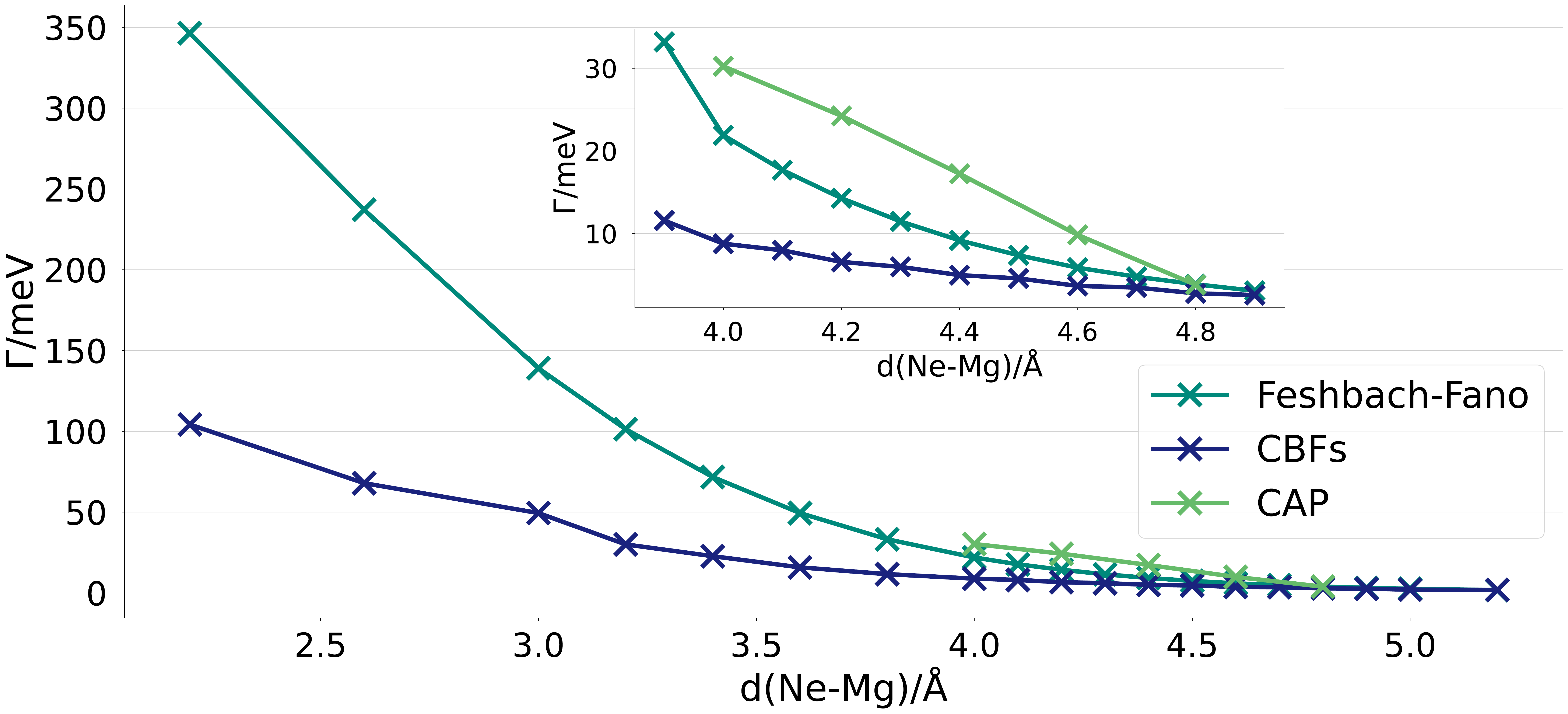}
\caption{Total decay width of Ne(2s$^{-1}$)Mg$^+$ as a function of the 
interatomic distance computed with CBF-EOM-CCSD/aug-cc-pCVTZ(5sp)+2s2p2d 
(this work), Feshbach-Fano-EOM-CCSD/aug-cc-pV5Z (this work), and 
CAP-EOM-CCSD/aug-cc-pVTZ (Ref. \onlinecite{Ghosh14}).}
\label{fig:nemg_w}
\end{figure}

The decay width of Ne(2s$^{-1}$)Mg$^+$ and its dependence on the 
interatomic distance has been computed several times.\cite{Averbukh04,
Averbukh05,Ghosh14} In Figure \ref{fig:nemg_w}, we report 
CAP-EOM-IP-CCSD results from Ref. \onlinecite{Ghosh14} alongside our 
CBF-EOM-IP-CCSD and Feshbach-Fano-EOM-CCSD results. For CBF-EOM-IP-CCSD, 
the basis set is aug-cc-pCVTZ(5sp)+2s2p2d; the complex exponents for 
the neon atom are the same that were used for Ne$_2$ and NeAr, while 
for the magnesium atom they are obtained by the procedure from Sec. 
\ref{section:theory_cbf}. Using the exponents for neon as starting 
point resulted in exponents of 0.187 and 0.030 for magnesium. Similar 
to NeAr, we use the same exponents for all distances (2.2--5.2 \AA) 
because the difference between the energy of the resonance and the 
target-state energies does not change substantially. 

CBF-EOM-IP-CCSD delivers the sum of the ICD and ETMD widths although 
this can be approximated as the ICD width given that the ETMD 
contribution is very small. At the equilibrium distance of the 
neutral ground state (4.4 \AA), we compute a width of 9.2 meV, 
which is lower than the CAP-EOM-IP-CCSD value of 17.25 meV reported 
by Ghosh \textit{et al.}\cite{Ghosh14} As concerns the distance 
dependence, CBF-EOM-IP CCSD yields $\Gamma(R) \sim R^{-5.11}$ 
meaning that the deviation from the expected result ($R^{-6}$) is 
somewhat larger than for NeAr. Contrary to what we saw for NeAr, 
here the agreement is better at larger distances. When the four 
points at the shortest distances are removed, we obtain $\Gamma(R) 
\sim R^{-6.03}$. A possible explanation of the deviation of 
$\Gamma(R)$ from $R^{-6}$ at short interatomic distances could 
be that orbital overlap starts playing a significant role, which 
would make it inappropriate to approximate the full width by 
the direct ICD contribution.\cite{Averbukh04} 

As concerns Feshbach-Fano-EOM-CCSD, Fig. \ref{fig:nemg_w} shows 
that this method yields larger widths than CBF-EOM-IP-CCSD at all 
distances considered, in particular at short distances. 
Moreover, the width does not follow a power law: in a log-log plot 
we do not obtain a line but rather a quadratic function. Notably, 
CAP-EOM-IP-CCSD predicts even larger widths at all interatomic 
distances reported in Ref. \onlinecite{Ghosh14} (4.0--4.8 \AA). 
However, the discrepancies between the methods decrease as the 
atoms are moved apart. 

\section{Hydrogen fluoride dimer} \label{sec:hf2}

\begin{figure}[ht]
\includegraphics[width=8cm,trim=110 100 80 60,clip]{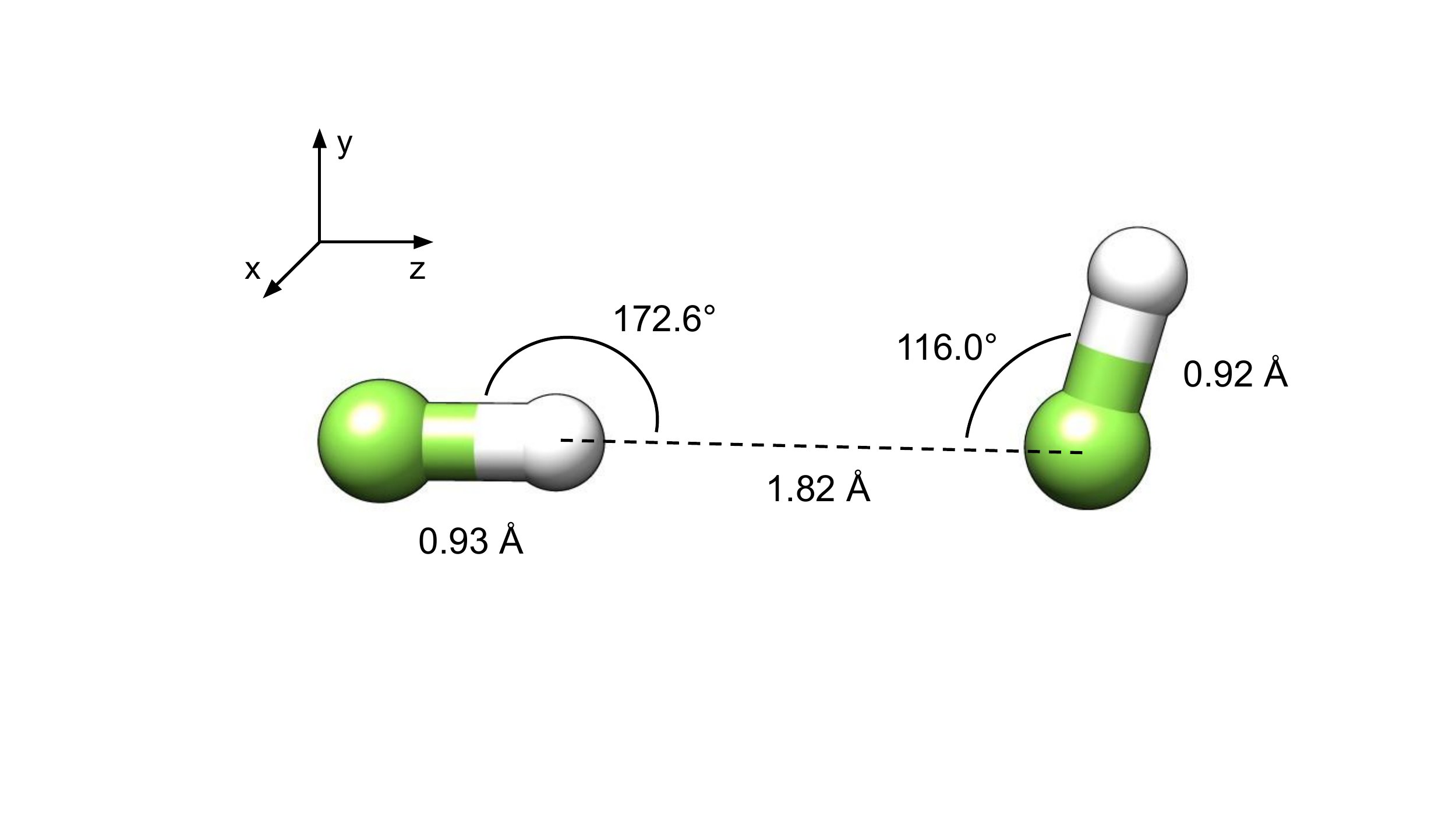}
\caption{Optimized structure of (HF)$_2$}
\label{fig:hf_geom}
\end{figure}

As an example of ICD in a molecular system, we computed the decay 
rate of (HF)$_2$. The optimized structure of this dimer is shown 
in Fig. \ref{fig:hf_geom}. This structure was taken from Ref. 
\onlinecite{Santra99} and re-optimized with RI-MP2/aug-cc-pVTZ; 
the coordinates are reported in the Supplementary Material. In 
(HF)$_2$, two distinct resonances result from ionizing the 2s 
orbitals of the fluorine atoms. They can be distinguished because 
the vacancies are localized on the fluorine atom of the molecule 
that acts as H-bond donor or the one that acts as H-bond acceptor. 
In this work, we focus on the lower-lying state in which the 
ionized fluorine atom is the one of the H-bond donor, which is 
shown on the left in Fig. \ref{fig:hf_geom}.

The inner-valence ionization energy is reported in Tab. 
\ref{tab:hf_w}, while the energies of the accessible target 
states can be found in Tab. \ref{tab:hf_dip}. Our results for 
the latter states are qualitatively in line with those from 
Ref. \onlinecite{Zobeley98} although the energies differ 
slightly. Notably, the energy of the ICD electron is only 
0.3 eV in (HF)$_2^+$, which is much smaller compared to the 
dimers examined in the previous sections. 

Our results for the decay width are reported in Tab. 
\ref{tab:hf_w}. These were computed using the 
aug-cc-pCVTZ(5sp)+2s2p2d basis, where the exponents of the 
complex-scaled shells are again obtained by means of the 
procedure described in Sec. \ref{section:theory_cbf} using 
the exponents for Ne$_2$ as starting point. The values are 
0.0915, 0.0057 and 0.985, 0.062 for H and F, respectively. 
As can be seen from Tab. \ref{tab:hf_w}, the resonance 
positions and widths obtained with CBF-EOM-IP-CCSD are in 
good agreement with CAP-CI and CAP-EOM-CCSD values available 
from the literature.\cite{Santra99,Ghosh14-2} For what 
concerns Feshbach-Fano-EOM-CCSD, however, the total width 
is one order of magnitude too small, which shows again that 
the method devised for Auger decay is not suitable to describe 
ICD.

\begin{table}[ht]
\caption{Ionization energy and decay width of the lowest 
resonance state of (HF)$_2^+$(2s$^{-1}$) computed with various 
electronic structure methods.}
\begin{ruledtabular}
\begin{tabular}{lcc} 
Method & Energy/eV & Width/meV \\  
\hline
\makecell[l]{CBF-EOM-CCSD/aug-cc-pCVTZ(5sp) \\ +2s2p2d 
(this work)} & 38.621 & 17.0 \\
\makecell[l]{Feshbach-Fano-EOM-CCSD/ \\ aug-cc-pV5Z (this work)} 
& 39.18 & 1.8 \\
CAP-CI/aug-cc-pVTZ \cite{Santra99} & 38.6 & 18.0 \\ 
CAP-EOM-CCSD/aug-cc-pVTZ \cite{Ghosh14-2} & 39.1 & 20.7 \\ 
\end{tabular}
\end{ruledtabular}
\label{tab:hf_w}
\end{table}

\begin{table}[h]
\caption{Double ionization energies of the states resulting 
from the decay of (HF)$_2^+$(2s$^{-1}$) computed with 
EOM-CCSD/aug-cc-pCVTZ(5sp). Subscripts $d$ and $a$ refer 
to the molecule that acts as H-bond donor and acceptor, 
respectively.}
\centering
\begin{ruledtabular}
{\begin{tabular}{lccc} 
ICD final state & Singlet/eV  & Triplet/eV \\ \hline
HF$_d$(2p$_x^{-1}$)HF$_a$(2p$_x^{-1}$) & 38.341 & 38.342 \\
HF$_d$(2p$_y^{-1}$)HF$_a$(3p$_x^{-1}$) & 38.345 & 38.346 \\
\end{tabular}}
\end{ruledtabular}
\label{tab:hf_dip}
\end{table}



\section{Discussion and conclusions} \label{sec:disc}

In this work, we investigated interatomic/intermolecular Coulombic 
decay (ICD) in clusters of atoms and molecules by means of the 
equation-of-motion coupled-cluster singles and doubles method 
combined with complex basis functions and, alternatively, 
Feshbach-Fano projection and a plane wave description for the 
emitted electron. 

In the three atomic dimers Ne$_2$, NeAr, and NeMg, the initially 
ionized orbital is always a neon 2s orbital. Consequently, the 
ionization energy needed to create the vacancy differs by less 
than 0.1 eV among the three dimers, whereas the ionized state 
of (HF)$_2$ is 10 eV lower in energy because the vacancy is in 
a fluorine 2s orbital.

The energy of the ICD electrons follows a different trend: It 
assumes values of 0.8 eV for Ne$_2$ and 0.3 eV for (HF)$_2$ but 
7 eV and 16 eV for NeAr and NeMg, respectively. In general, the 
energy of the ICD electron is higher when the two participating 
atoms or molecules are different, as the orbital energy differences 
are bigger in this case. For the same reason, only the inner-valence 
vacancies of NeAr and NeMg can relax via electron transfer mediated 
decay (ETMD). Interestingly, the ETMD electron has a lower energy 
than the ICD electron in NeAr, while it is larger in the case of NeMg. 

To describe ICD and ETMD, complex basis functions offer a framework 
in which the outgoing electron does not need to be described explicitly. 
Rather, the decay width is directly obtained from the imaginary part 
of the total energy. In general, CBF-EOM-IP-CCSD yields results that 
are in good agreement with previously reported results based on 
complex absorbing potentials and Fano-Feshbach projection combined 
with Stieltjes imaging.

However, we found that the performance of CBF-EOM-IP-CCSD is much 
better for NeAr and NeMg than for Ne$_2$ and (HF)$_2$. For the 
former two systems, the distance dependence of the computed decay 
widths is in excellent agreement with the expectation, that is, 
it is proportional to $R^{-6}$. In contrast, for Ne$_2$ the distance 
dependence of the width is less well described. Also, the width of 
Ne$_2$ is much more sensitive to changes in the complex-scaled 
exponents than that of NeAr. We relate both shortcomings to the 
ICD electron having less energy in Ne$_2$ as compared to NeAr and 
NeMg.

Compared to Auger decay, ICD is more difficult to describe with 
CBF methods. This is not only because the emitted electron is 
slower but also because ICD widths are typically smaller than 
Auger decay widths. At their equilibrium structures, the systems 
considered in this work all have widths between 6 and 17 meV, 
whereas Auger decay widths can be greater than 100 meV for 
first-row nuclei and even greater for heavier nuclei. From the 
analysis of the distance dependence of the widths of NeAr and 
NeMg, we conclude that values down to ca. 5 meV can be reliably 
computed with CBF-EOM-IP-CCSD. 

Trends in the ICD width among the clusters are influenced by 
the differences in the interatomic distances and the orbital 
energies. Notably, the ICD width is very sensitive to a change 
of the partner atom, while it is almost independent of the 
energy of the emitted electron, meaning a large decay width 
does not correspond to a fast electron. 

Finally, we note that the Feshbach-Fano approach combined with 
a plane-wave description for the emitted electron is not suitable 
to describe ICD. This approach was devised for the treatment 
of Auger decay where it delivers good results but is unable to 
capture the characteristic $R^{-6}$ distance dependence of the 
ICD width. We ascribe this failure to using plane waves as 
previously reported applications of the Feshbach-Fano approach 
to ICD relying on Stieltjes imaging yielded convincing results.

\section*{Supplementary Material}
Supplementary material is available: geometry of (HF)$_2$, exponents 
of Ne$_2$ as well as double ionization energies and decay widths for 
all dimers discussed in the article.

\begin{acknowledgments}
T.-C.J. gratefully acknowledges funding from the European Research 
Council (ERC) under the European Union’s Horizon 2020 research and 
innovation program (Grant Agreement No. 851766) and the KU Leuven 
internal funds (Grant C14/22/083). We thank Florian Matz and Anthuan 
Ferino-Pérez for insightful discussions.
\end{acknowledgments}

\section*{Author Declarations}

\subsection*{Conflict of Interest Statement}
The authors have no conflicts to disclose.

\subsection*{Author Contributions}
\textbf{Valentina Parravicini:} Conceptualization (equal); Methodology (lead); 
Investigation (lead); Writing - original draft (lead).
\textbf{Thomas-C. Jagau:} Conceptualization (equal); Methodology (supporting); 
Investigation (supporting); Writing - original draft (supporting); Writing 
review \& editing (lead); Funding acquisition (lead).

\section*{Data Availability}
The data that support the findings of this study are available within the 
article and its supplementary material.

\section*{References} 
\bibliography{bibliography}

\begin{thebibliography}{66}%
\makeatletter
\providecommand \@ifxundefined [1]{%
 \@ifx{#1\undefined}
}%
\providecommand \@ifnum [1]{%
 \ifnum #1\expandafter \@firstoftwo
 \else \expandafter \@secondoftwo
 \fi
}%
\providecommand \@ifx [1]{%
 \ifx #1\expandafter \@firstoftwo
 \else \expandafter \@secondoftwo
 \fi
}%
\providecommand \natexlab [1]{#1}%
\providecommand \enquote  [1]{``#1''}%
\providecommand \bibnamefont  [1]{#1}%
\providecommand \bibfnamefont [1]{#1}%
\providecommand \citenamefont [1]{#1}%
\providecommand \href@noop [0]{\@secondoftwo}%
\providecommand \href [0]{\begingroup \@sanitize@url \@href}%
\providecommand \@href[1]{\@@startlink{#1}\@@href}%
\providecommand \@@href[1]{\endgroup#1\@@endlink}%
\providecommand \@sanitize@url [0]{\catcode `\\12\catcode `\$12\catcode
  `\&12\catcode `\#12\catcode `\^12\catcode `\_12\catcode `\%12\relax}%
\providecommand \@@startlink[1]{}%
\providecommand \@@endlink[0]{}%
\providecommand \url  [0]{\begingroup\@sanitize@url \@url }%
\providecommand \@url [1]{\endgroup\@href {#1}{\urlprefix }}%
\providecommand \urlprefix  [0]{URL }%
\providecommand \Eprint [0]{\href }%
\providecommand \doibase [0]{http://dx.doi.org/}%
\providecommand \selectlanguage [0]{\@gobble}%
\providecommand \bibinfo  [0]{\@secondoftwo}%
\providecommand \bibfield  [0]{\@secondoftwo}%
\providecommand \translation [1]{[#1]}%
\providecommand \BibitemOpen [0]{}%
\providecommand \bibitemStop [0]{}%
\providecommand \bibitemNoStop [0]{.\EOS\space}%
\providecommand \EOS [0]{\spacefactor3000\relax}%
\providecommand \BibitemShut  [1]{\csname bibitem#1\endcsname}%
\let\auto@bib@innerbib\@empty
\bibitem [{\citenamefont {Cederbaum}, \citenamefont {Zobeley},\ and\
  \citenamefont {Tarantelli}(1997)}]{Cederbaum97}%
  \BibitemOpen
  \bibfield  {author} {\bibinfo {author} {\bibfnamefont {L.~S.}\ \bibnamefont
  {Cederbaum}}, \bibinfo {author} {\bibfnamefont {J.}~\bibnamefont {Zobeley}},
  \ and\ \bibinfo {author} {\bibfnamefont {F.}~\bibnamefont {Tarantelli}},\
  }\bibfield  {title} {\enquote {\bibinfo {title} {Giant intermolecular decay
  and fragmentation of clusters},}\ }\href {\doibase
  10.1103/PhysRevLett.79.4778} {\bibfield  {journal} {\bibinfo  {journal}
  {Phys. Rev. Lett.}\ }\textbf {\bibinfo {volume} {79}},\ \bibinfo {pages}
  {4778–4781} (\bibinfo {year} {1997})}\BibitemShut {NoStop}%
\bibitem [{\citenamefont {Jahnke}\ \emph {et~al.}(2020)\citenamefont {Jahnke},
  \citenamefont {Hergenhahn}, \citenamefont {Winter}, \citenamefont {Dörner},
  \citenamefont {Frühling}, \citenamefont {Demekhin}, \citenamefont
  {Gokhberg}, \citenamefont {Cederbaum}, \citenamefont {Ehresmann},
  \citenamefont {Knie},\ and\ \citenamefont {Dreuw}}]{Jahnke20}%
  \BibitemOpen
  \bibfield  {author} {\bibinfo {author} {\bibfnamefont {T.}~\bibnamefont
  {Jahnke}}, \bibinfo {author} {\bibfnamefont {U.}~\bibnamefont {Hergenhahn}},
  \bibinfo {author} {\bibfnamefont {B.}~\bibnamefont {Winter}}, \bibinfo
  {author} {\bibfnamefont {R.}~\bibnamefont {Dörner}}, \bibinfo {author}
  {\bibfnamefont {U.}~\bibnamefont {Frühling}}, \bibinfo {author}
  {\bibfnamefont {P.~V.}\ \bibnamefont {Demekhin}}, \bibinfo {author}
  {\bibfnamefont {K.}~\bibnamefont {Gokhberg}}, \bibinfo {author}
  {\bibfnamefont {L.~S.}\ \bibnamefont {Cederbaum}}, \bibinfo {author}
  {\bibfnamefont {A.}~\bibnamefont {Ehresmann}}, \bibinfo {author}
  {\bibfnamefont {A.}~\bibnamefont {Knie}}, \ and\ \bibinfo {author}
  {\bibfnamefont {A.}~\bibnamefont {Dreuw}},\ }\bibfield  {title} {\enquote
  {\bibinfo {title} {Interatomic and intermolecular {Coulombic} decay},}\
  }\href {\doibase 10.1021/acs.chemrev.0c00106} {\bibfield  {journal} {\bibinfo
   {journal} {Chemical Reviews}\ }\textbf {\bibinfo {volume} {120}},\ \bibinfo
  {pages} {11295–11369} (\bibinfo {year} {2020})}\BibitemShut {NoStop}%
\bibitem [{\citenamefont {Meitner}(1922)}]{Meitner22}%
  \BibitemOpen
  \bibfield  {author} {\bibinfo {author} {\bibfnamefont {L.}~\bibnamefont
  {Meitner}},\ }\bibfield  {title} {\enquote {\bibinfo {title} {Über die
  $\beta$-{S}trahl-{S}pektra und ihren {Z}usammenhang mit der
  $\gamma$-{S}trahlung},}\ }\href {\doibase 10.1007/BF01328399} {\bibfield
  {journal} {\bibinfo  {journal} {Z. Phys.}\ }\textbf {\bibinfo {volume}
  {11}},\ \bibinfo {pages} {35–54} (\bibinfo {year} {1922})}\BibitemShut
  {NoStop}%
\bibitem [{\citenamefont {{Auger}}(1923)}]{Auger23}%
  \BibitemOpen
  \bibfield  {author} {\bibinfo {author} {\bibnamefont {{Auger}}},\ }\bibfield
  {title} {\enquote {\bibinfo {title} {Sur les rayons beta secondaires produits
  dans un gaz par des rayons x},}\ }\href@noop {} {\bibfield  {journal}
  {\bibinfo  {journal} {C.R.A.S.}\ }\textbf {\bibinfo {volume} {177}},\
  \bibinfo {pages} {169–171} (\bibinfo {year} {1923})}\BibitemShut {NoStop}%
\bibitem [{\citenamefont {Santra}, \citenamefont {Zobeley},\ and\ \citenamefont
  {Cederbaum}(2001)}]{Santra01-2}%
  \BibitemOpen
  \bibfield  {author} {\bibinfo {author} {\bibfnamefont {R.}~\bibnamefont
  {Santra}}, \bibinfo {author} {\bibfnamefont {J.}~\bibnamefont {Zobeley}}, \
  and\ \bibinfo {author} {\bibfnamefont {L.}~\bibnamefont {Cederbaum}},\
  }\bibfield  {title} {\enquote {\bibinfo {title} {Electronic decay of valence
  holes in clusters and condensed matter},}\ }\href {\doibase
  10.1103/PhysRevB.64.245104} {\bibfield  {journal} {\bibinfo  {journal} {Phys.
  Rev. B}\ }\textbf {\bibinfo {volume} {64}},\ \bibinfo {pages} {245104}
  (\bibinfo {year} {2001})}\BibitemShut {NoStop}%
\bibitem [{\citenamefont {Thomas}\ \emph {et~al.}(2002)\citenamefont {Thomas},
  \citenamefont {Miron}, \citenamefont {Wiesner}, \citenamefont {Morin},
  \citenamefont {Carroll},\ and\ \citenamefont {S\ae{}thre}}]{Thomas02}%
  \BibitemOpen
  \bibfield  {author} {\bibinfo {author} {\bibfnamefont {T.~D.}\ \bibnamefont
  {Thomas}}, \bibinfo {author} {\bibfnamefont {C.}~\bibnamefont {Miron}},
  \bibinfo {author} {\bibfnamefont {K.}~\bibnamefont {Wiesner}}, \bibinfo
  {author} {\bibfnamefont {P.}~\bibnamefont {Morin}}, \bibinfo {author}
  {\bibfnamefont {T.~X.}\ \bibnamefont {Carroll}}, \ and\ \bibinfo {author}
  {\bibfnamefont {L.~J.}\ \bibnamefont {S\ae{}thre}},\ }\bibfield  {title}
  {\enquote {\bibinfo {title} {Anomalous natural linewidth in the $2p$
  photoelectron spectrum of {SiF$_4$}},}\ }\href {\doibase
  10.1103/PhysRevLett.89.223001} {\bibfield  {journal} {\bibinfo  {journal}
  {Phys. Rev. Lett.}\ }\textbf {\bibinfo {volume} {89}},\ \bibinfo {pages}
  {223001} (\bibinfo {year} {2002})}\BibitemShut {NoStop}%
\bibitem [{\citenamefont {Averbukh}, \citenamefont {Müller},\ and\
  \citenamefont {Cederbaum}(2004)}]{Averbukh04}%
  \BibitemOpen
  \bibfield  {author} {\bibinfo {author} {\bibfnamefont {V.}~\bibnamefont
  {Averbukh}}, \bibinfo {author} {\bibfnamefont {I.~B.}\ \bibnamefont
  {Müller}}, \ and\ \bibinfo {author} {\bibfnamefont {L.~S.}\ \bibnamefont
  {Cederbaum}},\ }\bibfield  {title} {\enquote {\bibinfo {title} {Mechanism of
  interatomic {Coulombic} decay in clusters},}\ }\href {\doibase
  10.1103/PhysRevLett.93.263002} {\bibfield  {journal} {\bibinfo  {journal}
  {Phys. Rev. Lett.}\ }\textbf {\bibinfo {volume} {93}},\ \bibinfo {pages}
  {263002} (\bibinfo {year} {2004})}\BibitemShut {NoStop}%
\bibitem [{\citenamefont {Miteva}\ \emph {et~al.}(2017)\citenamefont {Miteva},
  \citenamefont {Kazandjian}, \citenamefont {Kolorenč}, \citenamefont
  {Votavová},\ and\ \citenamefont {Sisourat}}]{Miteva17}%
  \BibitemOpen
  \bibfield  {author} {\bibinfo {author} {\bibfnamefont {T.}~\bibnamefont
  {Miteva}}, \bibinfo {author} {\bibfnamefont {S.}~\bibnamefont {Kazandjian}},
  \bibinfo {author} {\bibfnamefont {P.}~\bibnamefont {Kolorenč}}, \bibinfo
  {author} {\bibfnamefont {P.}~\bibnamefont {Votavová}}, \ and\ \bibinfo
  {author} {\bibfnamefont {N.}~\bibnamefont {Sisourat}},\ }\bibfield  {title}
  {\enquote {\bibinfo {title} {Interatomic {Coulombic} decay mediated by
  ultrafast superexchange energy transfer},}\ }\href {\doibase
  10.1103/PhysRevLett.119.083403} {\bibfield  {journal} {\bibinfo  {journal}
  {Phys. Rev. Lett.}\ }\textbf {\bibinfo {volume} {119}},\ \bibinfo {pages}
  {083403} (\bibinfo {year} {2017})}\BibitemShut {NoStop}%
\bibitem [{\citenamefont {Votavová}\ \emph {et~al.}(2019)\citenamefont
  {Votavová}, \citenamefont {Miteva}, \citenamefont {Engin}, \citenamefont
  {Kazandjian}, \citenamefont {Kolorenč},\ and\ \citenamefont
  {Sisourat}}]{Votavova19}%
  \BibitemOpen
  \bibfield  {author} {\bibinfo {author} {\bibfnamefont {P.}~\bibnamefont
  {Votavová}}, \bibinfo {author} {\bibfnamefont {T.}~\bibnamefont {Miteva}},
  \bibinfo {author} {\bibfnamefont {S.}~\bibnamefont {Engin}}, \bibinfo
  {author} {\bibfnamefont {S.}~\bibnamefont {Kazandjian}}, \bibinfo {author}
  {\bibfnamefont {P.}~\bibnamefont {Kolorenč}}, \ and\ \bibinfo {author}
  {\bibfnamefont {N.}~\bibnamefont {Sisourat}},\ }\bibfield  {title} {\enquote
  {\bibinfo {title} {Mechanism of superexchange interatomic {Coulombic} decay
  in rare-gas clusters},}\ }\href {\doibase 10.1103/PhysRevA.100.022706}
  {\bibfield  {journal} {\bibinfo  {journal} {Phys. Rev. A}\ }\textbf {\bibinfo
  {volume} {100}},\ \bibinfo {pages} {022706} (\bibinfo {year}
  {2019})}\BibitemShut {NoStop}%
\bibitem [{\citenamefont {Öhrwall}\ \emph {et~al.}(2004)\citenamefont
  {Öhrwall}, \citenamefont {Tchaplyguine}, \citenamefont {Lundwall},
  \citenamefont {Feifel}, \citenamefont {Bergersen}, \citenamefont {Rander},
  \citenamefont {Lindblad}, \citenamefont {Schulz}, \citenamefont {Peredkov},
  \citenamefont {Barth}, \citenamefont {Marburger}, \citenamefont {Hergenhahn},
  \citenamefont {Svensson},\ and\ \citenamefont {Björneholm}}]{Ohrwall04}%
  \BibitemOpen
  \bibfield  {author} {\bibinfo {author} {\bibfnamefont {G.}~\bibnamefont
  {Öhrwall}}, \bibinfo {author} {\bibfnamefont {M.}~\bibnamefont
  {Tchaplyguine}}, \bibinfo {author} {\bibfnamefont {M.}~\bibnamefont
  {Lundwall}}, \bibinfo {author} {\bibfnamefont {R.}~\bibnamefont {Feifel}},
  \bibinfo {author} {\bibfnamefont {H.}~\bibnamefont {Bergersen}}, \bibinfo
  {author} {\bibfnamefont {T.}~\bibnamefont {Rander}}, \bibinfo {author}
  {\bibfnamefont {A.}~\bibnamefont {Lindblad}}, \bibinfo {author}
  {\bibfnamefont {J.}~\bibnamefont {Schulz}}, \bibinfo {author} {\bibfnamefont
  {S.}~\bibnamefont {Peredkov}}, \bibinfo {author} {\bibfnamefont
  {S.}~\bibnamefont {Barth}}, \bibinfo {author} {\bibfnamefont
  {S.}~\bibnamefont {Marburger}}, \bibinfo {author} {\bibfnamefont
  {U.}~\bibnamefont {Hergenhahn}}, \bibinfo {author} {\bibfnamefont
  {S.}~\bibnamefont {Svensson}}, \ and\ \bibinfo {author} {\bibfnamefont
  {O.}~\bibnamefont {Björneholm}},\ }\bibfield  {title} {\enquote {\bibinfo
  {title} {Femtosecond interatomic {Coulombic} decay in free neon clusters:
  Large lifetime differences between surface and bulk},}\ }\href {\doibase
  10.1103/PhysRevLett.93.173401} {\bibfield  {journal} {\bibinfo  {journal}
  {Phys. Rev. Lett.}\ }\textbf {\bibinfo {volume} {93}},\ \bibinfo {pages}
  {173401} (\bibinfo {year} {2004})}\BibitemShut {NoStop}%
\bibitem [{\citenamefont {Barth}\ \emph {et~al.}(2006)\citenamefont {Barth},
  \citenamefont {Marburger}, \citenamefont {Kugeler}, \citenamefont {Ulrich},
  \citenamefont {Joshi}, \citenamefont {Bradshaw},\ and\ \citenamefont
  {Hergenhahn}}]{Barth06}%
  \BibitemOpen
  \bibfield  {author} {\bibinfo {author} {\bibfnamefont {S.}~\bibnamefont
  {Barth}}, \bibinfo {author} {\bibfnamefont {S.}~\bibnamefont {Marburger}},
  \bibinfo {author} {\bibfnamefont {O.}~\bibnamefont {Kugeler}}, \bibinfo
  {author} {\bibfnamefont {V.}~\bibnamefont {Ulrich}}, \bibinfo {author}
  {\bibfnamefont {S.}~\bibnamefont {Joshi}}, \bibinfo {author} {\bibfnamefont
  {A.~M.}\ \bibnamefont {Bradshaw}}, \ and\ \bibinfo {author} {\bibfnamefont
  {U.}~\bibnamefont {Hergenhahn}},\ }\bibfield  {title} {\enquote {\bibinfo
  {title} {The efficiency of interatomic {Coulombic} decay in {Ne} clusters},}\
  }\href {\doibase 10.1016/j.chemphys.2006.06.035} {\bibfield  {journal}
  {\bibinfo  {journal} {Chem. Phys.}\ }\bibinfo {series} {Electron Correlation
  and Multimode Dynamics in Molecules},\ \textbf {\bibinfo {volume} {329}},\
  \bibinfo {pages} {246–250} (\bibinfo {year} {2006})}\BibitemShut {NoStop}%
\bibitem [{\citenamefont {Ghosh}\ and\ \citenamefont {Vaval}(2014)}]{Ghosh14}%
  \BibitemOpen
  \bibfield  {author} {\bibinfo {author} {\bibfnamefont {A.}~\bibnamefont
  {Ghosh}}\ and\ \bibinfo {author} {\bibfnamefont {N.}~\bibnamefont {Vaval}},\
  }\bibfield  {title} {\enquote {\bibinfo {title} {Geometry-dependent lifetime
  of interatomic {Coulombic} decay using equation-of-motion coupled cluster
  method},}\ }\href {\doibase 10.1063/1.4903827} {\bibfield  {journal}
  {\bibinfo  {journal} {J. Chem. Phys.}\ }\textbf {\bibinfo {volume} {141}},\
  \bibinfo {pages} {234108} (\bibinfo {year} {2014})}\BibitemShut {NoStop}%
\bibitem [{\citenamefont {Fasshauer}(2016)}]{Fasshauer16}%
  \BibitemOpen
  \bibfield  {author} {\bibinfo {author} {\bibfnamefont {E.}~\bibnamefont
  {Fasshauer}},\ }\bibfield  {title} {\enquote {\bibinfo {title} {Non-nearest
  neighbour {ICD} in clusters},}\ }\href {\doibase
  10.1088/1367-2630/18/4/043028} {\bibfield  {journal} {\bibinfo  {journal}
  {New J. Phys.}\ }\textbf {\bibinfo {volume} {18}},\ \bibinfo {pages} {043028}
  (\bibinfo {year} {2016})}\BibitemShut {NoStop}%
\bibitem [{\citenamefont {Kumar}, \citenamefont {Ghosh},\ and\ \citenamefont
  {Vaval}(2022)}]{Kumar22}%
  \BibitemOpen
  \bibfield  {author} {\bibinfo {author} {\bibfnamefont {R.}~\bibnamefont
  {Kumar}}, \bibinfo {author} {\bibfnamefont {A.}~\bibnamefont {Ghosh}}, \ and\
  \bibinfo {author} {\bibfnamefont {N.}~\bibnamefont {Vaval}},\ }\bibfield
  {title} {\enquote {\bibinfo {title} {Decay processes in cationic alkali
  metals in microsolvated clusters: A complex absorbing potential based
  equation-of-motion coupled cluster investigation},}\ }\href {\doibase
  10.1021/acs.jctc.1c01036} {\bibfield  {journal} {\bibinfo  {journal} {J.
  Chem. Theory Comput.}\ }\textbf {\bibinfo {volume} {18}},\ \bibinfo {pages}
  {807–816} (\bibinfo {year} {2022})}\BibitemShut {NoStop}%
\bibitem [{\citenamefont {Zobeley}, \citenamefont {Santra},\ and\ \citenamefont
  {Cederbaum}(2001)}]{Zobeley01}%
  \BibitemOpen
  \bibfield  {author} {\bibinfo {author} {\bibfnamefont {J.}~\bibnamefont
  {Zobeley}}, \bibinfo {author} {\bibfnamefont {R.}~\bibnamefont {Santra}}, \
  and\ \bibinfo {author} {\bibfnamefont {L.~S.}\ \bibnamefont {Cederbaum}},\
  }\bibfield  {title} {\enquote {\bibinfo {title} {Electronic decay in weakly
  bound heteroclusters: Energy transfer versus electron transfer},}\ }\href
  {\doibase 10.1063/1.1395555} {\bibfield  {journal} {\bibinfo  {journal} {J.
  Chem. Phys.}\ }\textbf {\bibinfo {volume} {115}},\ \bibinfo {pages}
  {5076–5088} (\bibinfo {year} {2001})}\BibitemShut {NoStop}%
\bibitem [{\citenamefont {Moiseyev}(2011)}]{Moiseyev11}%
  \BibitemOpen
  \bibfield  {author} {\bibinfo {author} {\bibfnamefont {N.}~\bibnamefont
  {Moiseyev}},\ }\href@noop {} {\emph {\bibinfo {title} {Non-{Hermitian}
  Quantum Mechanics}}}\ (\bibinfo  {publisher} {Cambridge University Press},\
  \bibinfo {year} {2011})\BibitemShut {NoStop}%
\bibitem [{\citenamefont {Jagau}, \citenamefont {Bravaya},\ and\ \citenamefont
  {Krylov}(2017)}]{Jagau17}%
  \BibitemOpen
  \bibfield  {author} {\bibinfo {author} {\bibfnamefont {T.-C.}\ \bibnamefont
  {Jagau}}, \bibinfo {author} {\bibfnamefont {K.~B.}\ \bibnamefont {Bravaya}},
  \ and\ \bibinfo {author} {\bibfnamefont {A.~I.}\ \bibnamefont {Krylov}},\
  }\bibfield  {title} {\enquote {\bibinfo {title} {Extending quantum chemistry
  of bound states to electronic resonances},}\ }\href {\doibase
  10.1146/annurev-physchem-052516-050622} {\bibfield  {journal} {\bibinfo
  {journal} {Annu. Rev. Phys. Chem.}\ }\textbf {\bibinfo {volume} {68}},\
  \bibinfo {pages} {525–553} (\bibinfo {year} {2017})}\BibitemShut {NoStop}%
\bibitem [{\citenamefont {Jagau}(2022)}]{Jagau22}%
  \BibitemOpen
  \bibfield  {author} {\bibinfo {author} {\bibfnamefont {T.-C.}\ \bibnamefont
  {Jagau}},\ }\bibfield  {title} {\enquote {\bibinfo {title} {Theory of
  electronic resonances: fundamental aspects and recent advances},}\ }\href
  {\doibase 10.1039/D1CC07090H} {\bibfield  {journal} {\bibinfo  {journal}
  {Chem. Commun.}\ }\textbf {\bibinfo {volume} {58}},\ \bibinfo {pages}
  {5205–5224} (\bibinfo {year} {2022})}\BibitemShut {NoStop}%
\bibitem [{\citenamefont {{Fano}}(1961)}]{Fano61}%
  \BibitemOpen
  \bibfield  {author} {\bibinfo {author} {\bibfnamefont {U.}~\bibnamefont
  {{Fano}}},\ }\bibfield  {title} {\enquote {\bibinfo {title} {Effects of
  configuration interaction on intensities and phase shifts},}\ }\href
  {\doibase 10.1103/PhysRev.124.1866} {\bibfield  {journal} {\bibinfo
  {journal} {Phys. Rev.}\ }\textbf {\bibinfo {volume} {124}},\ \bibinfo {pages}
  {1866–1878} (\bibinfo {year} {1961})}\BibitemShut {NoStop}%
\bibitem [{\citenamefont {Feshbach}(1962)}]{Feshbach62}%
  \BibitemOpen
  \bibfield  {author} {\bibinfo {author} {\bibfnamefont {H.}~\bibnamefont
  {Feshbach}},\ }\bibfield  {title} {\enquote {\bibinfo {title} {A unified
  theory of nuclear reactions. {II}},}\ }\href {\doibase
  10.1016/0003-4916(62)90221-X} {\bibfield  {journal} {\bibinfo  {journal}
  {Ann. Phys.}\ }\textbf {\bibinfo {volume} {19}},\ \bibinfo {pages}
  {287–313} (\bibinfo {year} {1962})}\BibitemShut {NoStop}%
\bibitem [{\citenamefont {Langhoff}\ and\ \citenamefont
  {Corcoran}(1974)}]{Langhoff03}%
  \BibitemOpen
  \bibfield  {author} {\bibinfo {author} {\bibfnamefont {P.~W.}\ \bibnamefont
  {Langhoff}}\ and\ \bibinfo {author} {\bibfnamefont {C.~T.}\ \bibnamefont
  {Corcoran}},\ }\bibfield  {title} {\enquote {\bibinfo {title} {{Stieltjes}
  imaging of photoabsorption and dispersion profiles},}\ }\href {\doibase
  10.1063/1.1681616} {\bibfield  {journal} {\bibinfo  {journal} {J. Chem.
  Phys.}\ }\textbf {\bibinfo {volume} {61}},\ \bibinfo {pages} {146--159}
  (\bibinfo {year} {1974})}\BibitemShut {NoStop}%
\bibitem [{\citenamefont {Averbukh}\ and\ \citenamefont
  {Cederbaum}(2005)}]{Averbukh05}%
  \BibitemOpen
  \bibfield  {author} {\bibinfo {author} {\bibfnamefont {V.}~\bibnamefont
  {Averbukh}}\ and\ \bibinfo {author} {\bibfnamefont {L.~S.}\ \bibnamefont
  {Cederbaum}},\ }\bibfield  {title} {\enquote {\bibinfo {title} {Ab initio
  calculation of interatomic decay rates by a combination of the {Fano} ansatz,
  {Green}’s-function methods, and the {Stieltjes} imaging technique},}\
  }\href {\doibase 10.1063/1.2126976} {\bibfield  {journal} {\bibinfo
  {journal} {J. Chem. Phys.}\ }\textbf {\bibinfo {volume} {123}},\ \bibinfo
  {pages} {204107} (\bibinfo {year} {2005})}\BibitemShut {NoStop}%
\bibitem [{\citenamefont {Kolorenč}\ and\ \citenamefont
  {Averbukh}(2020)}]{Kolorenc20}%
  \BibitemOpen
  \bibfield  {author} {\bibinfo {author} {\bibfnamefont {P.}~\bibnamefont
  {Kolorenč}}\ and\ \bibinfo {author} {\bibfnamefont {V.}~\bibnamefont
  {Averbukh}},\ }\bibfield  {title} {\enquote {\bibinfo {title}
  {{Fano}-{ADC}(2,2) method for electronic decay rates},}\ }\href {\doibase
  10.1063/5.0007912} {\bibfield  {journal} {\bibinfo  {journal} {J. Chem.
  Phys.}\ }\textbf {\bibinfo {volume} {152}},\ \bibinfo {pages} {214107}
  (\bibinfo {year} {2020})}\BibitemShut {NoStop}%
\bibitem [{\citenamefont {Miteva}, \citenamefont {Kazandjian},\ and\
  \citenamefont {Sisourat}(2017)}]{Miteva17-2}%
  \BibitemOpen
  \bibfield  {author} {\bibinfo {author} {\bibfnamefont {T.}~\bibnamefont
  {Miteva}}, \bibinfo {author} {\bibfnamefont {S.}~\bibnamefont {Kazandjian}},
  \ and\ \bibinfo {author} {\bibfnamefont {N.}~\bibnamefont {Sisourat}},\
  }\bibfield  {title} {\enquote {\bibinfo {title} {On the computations of decay
  widths of {Fano} resonances},}\ }\href {\doibase
  10.1016/j.chemphys.2016.08.014} {\bibfield  {journal} {\bibinfo  {journal}
  {Chem. Phys.}\ }\textbf {\bibinfo {volume} {482}},\ \bibinfo {pages}
  {208–215} (\bibinfo {year} {2017})}\BibitemShut {NoStop}%
\bibitem [{\citenamefont {Sisourat}\ \emph {et~al.}(2017)\citenamefont
  {Sisourat}, \citenamefont {Engin}, \citenamefont {Gorfinkiel}, \citenamefont
  {Kazandjian}, \citenamefont {Kolorenč},\ and\ \citenamefont
  {Miteva}}]{Sisourat17}%
  \BibitemOpen
  \bibfield  {author} {\bibinfo {author} {\bibfnamefont {N.}~\bibnamefont
  {Sisourat}}, \bibinfo {author} {\bibfnamefont {S.}~\bibnamefont {Engin}},
  \bibinfo {author} {\bibfnamefont {J.~D.}\ \bibnamefont {Gorfinkiel}},
  \bibinfo {author} {\bibfnamefont {S.}~\bibnamefont {Kazandjian}}, \bibinfo
  {author} {\bibfnamefont {P.}~\bibnamefont {Kolorenč}}, \ and\ \bibinfo
  {author} {\bibfnamefont {T.}~\bibnamefont {Miteva}},\ }\bibfield  {title}
  {\enquote {\bibinfo {title} {On the computations of interatomic {Coulombic}
  decay widths with {R}-matrix method},}\ }\href {\doibase 10.1063/1.4989538}
  {\bibfield  {journal} {\bibinfo  {journal} {J. Chem. Phys.}\ }\textbf
  {\bibinfo {volume} {146}},\ \bibinfo {pages} {244109} (\bibinfo {year}
  {2017})}\BibitemShut {NoStop}%
\bibitem [{\citenamefont {Santra}\ and\ \citenamefont
  {Cederbaum}(2001)}]{Santra01}%
  \BibitemOpen
  \bibfield  {author} {\bibinfo {author} {\bibfnamefont {R.}~\bibnamefont
  {Santra}}\ and\ \bibinfo {author} {\bibfnamefont {L.~S.}\ \bibnamefont
  {Cederbaum}},\ }\bibfield  {title} {\enquote {\bibinfo {title} {An efficient
  combination of computational techniques for investigating electronic
  resonance states in molecules},}\ }\href {\doibase 10.1063/1.1405117}
  {\bibfield  {journal} {\bibinfo  {journal} {J. Chem. Phys.}\ }\textbf
  {\bibinfo {volume} {115}},\ \bibinfo {pages} {6853–6861} (\bibinfo {year}
  {2001})}\BibitemShut {NoStop}%
\bibitem [{\citenamefont {Vaval}\ and\ \citenamefont
  {Cederbaum}(2007)}]{Vaval07}%
  \BibitemOpen
  \bibfield  {author} {\bibinfo {author} {\bibfnamefont {N.}~\bibnamefont
  {Vaval}}\ and\ \bibinfo {author} {\bibfnamefont {L.~S.}\ \bibnamefont
  {Cederbaum}},\ }\bibfield  {title} {\enquote {\bibinfo {title} {Ab initio
  lifetimes in the interatomic {Coulombic} decay of neon clusters computed with
  propagators},}\ }\href {\doibase 10.1063/1.2723117} {\bibfield  {journal}
  {\bibinfo  {journal} {J. Chem. Phys.}\ }\textbf {\bibinfo {volume} {126}},\
  \bibinfo {pages} {164110} (\bibinfo {year} {2007})}\BibitemShut {NoStop}%
\bibitem [{\citenamefont {Ghosh}, \citenamefont {Pal},\ and\ \citenamefont
  {Vaval}(2013)}]{Ghosh13}%
  \BibitemOpen
  \bibfield  {author} {\bibinfo {author} {\bibfnamefont {A.}~\bibnamefont
  {Ghosh}}, \bibinfo {author} {\bibfnamefont {S.}~\bibnamefont {Pal}}, \ and\
  \bibinfo {author} {\bibfnamefont {N.}~\bibnamefont {Vaval}},\ }\bibfield
  {title} {\enquote {\bibinfo {title} {Study of interatomic {Coulombic} decay
  of {N}e({H$_2$O})$_n$ (n = 1,3) clusters using equation-of-motion
  coupled-cluster method},}\ }\href {\doibase 10.1063/1.4817966} {\bibfield
  {journal} {\bibinfo  {journal} {J. Chem. Phys.}\ }\textbf {\bibinfo {volume}
  {139}},\ \bibinfo {pages} {064112} (\bibinfo {year} {2013})}\BibitemShut
  {NoStop}%
\bibitem [{\citenamefont {Skomorowski}\ and\ \citenamefont
  {Krylov}(2021{\natexlab{a}})}]{Skomorowski21}%
  \BibitemOpen
  \bibfield  {author} {\bibinfo {author} {\bibfnamefont {W.}~\bibnamefont
  {Skomorowski}}\ and\ \bibinfo {author} {\bibfnamefont {A.~I.}\ \bibnamefont
  {Krylov}},\ }\bibfield  {title} {\enquote {\bibinfo {title} {Feshbach-{Fano}
  approach for calculation of {Auger} decay rates using equation-of-motion
  coupled-cluster wave functions. {I}. {Theory} and implementation},}\ }\href
  {\doibase 10.1063/5.0036976} {\bibfield  {journal} {\bibinfo  {journal} {J.
  Chem. Phys.}\ }\textbf {\bibinfo {volume} {154}},\ \bibinfo {pages} {084124}
  (\bibinfo {year} {2021}{\natexlab{a}})}\BibitemShut {NoStop}%
\bibitem [{\citenamefont {McCurdy}\ and\ \citenamefont
  {Rescigno}(1978)}]{McCurdy78}%
  \BibitemOpen
  \bibfield  {author} {\bibinfo {author} {\bibfnamefont {C.~W.}\ \bibnamefont
  {McCurdy}}\ and\ \bibinfo {author} {\bibfnamefont {T.~N.}\ \bibnamefont
  {Rescigno}},\ }\bibfield  {title} {\enquote {\bibinfo {title} {Extension of
  the method of complex basis functions to molecular resonances},}\ }\href
  {\doibase 10.1103/PhysRevLett.41.1364} {\bibfield  {journal} {\bibinfo
  {journal} {Phys. Rev. Lett.}\ }\textbf {\bibinfo {volume} {41}},\ \bibinfo
  {pages} {1364–1368} (\bibinfo {year} {1978})}\BibitemShut {NoStop}%
\bibitem [{\citenamefont {Stanton}\ and\ \citenamefont
  {Bartlett}(1993)}]{Stanton93}%
  \BibitemOpen
  \bibfield  {author} {\bibinfo {author} {\bibfnamefont {J.~F.}\ \bibnamefont
  {Stanton}}\ and\ \bibinfo {author} {\bibfnamefont {R.~J.}\ \bibnamefont
  {Bartlett}},\ }\bibfield  {title} {\enquote {\bibinfo {title} {The equation
  of motion coupled‐cluster method. a systematic biorthogonal approach to
  molecular excitation energies, transition probabilities, and excited state
  properties},}\ }\href {\doibase 10.1063/1.464746} {\bibfield  {journal}
  {\bibinfo  {journal} {J. Chem. Phys.}\ }\textbf {\bibinfo {volume} {98}},\
  \bibinfo {pages} {7029–7039} (\bibinfo {year} {1993})}\BibitemShut
  {NoStop}%
\bibitem [{\citenamefont {Sneskov}\ and\ \citenamefont
  {Christiansen}(2012)}]{Sneskov12}%
  \BibitemOpen
  \bibfield  {author} {\bibinfo {author} {\bibfnamefont {K.}~\bibnamefont
  {Sneskov}}\ and\ \bibinfo {author} {\bibfnamefont {O.}~\bibnamefont
  {Christiansen}},\ }\bibfield  {title} {\enquote {\bibinfo {title} {Excited
  state coupled cluster methods},}\ }\href {\doibase 10.1002/wcms.99}
  {\bibfield  {journal} {\bibinfo  {journal} {WIREs: Comp. Mol. Sci.}\ }\textbf
  {\bibinfo {volume} {2}},\ \bibinfo {pages} {566–584} (\bibinfo {year}
  {2012})}\BibitemShut {NoStop}%
\bibitem [{\citenamefont {Stanton}\ and\ \citenamefont
  {Gauss}(1994)}]{Stanton94}%
  \BibitemOpen
  \bibfield  {author} {\bibinfo {author} {\bibfnamefont {J.~F.}\ \bibnamefont
  {Stanton}}\ and\ \bibinfo {author} {\bibfnamefont {J.}~\bibnamefont
  {Gauss}},\ }\bibfield  {title} {\enquote {\bibinfo {title} {Analytic energy
  derivatives for ionized states described by the equation-of-motion coupled
  cluster method},}\ }\href {\doibase 10.1063/1.468022} {\bibfield  {journal}
  {\bibinfo  {journal} {J. Chem. Phys.}\ }\textbf {\bibinfo {volume} {101}},\
  \bibinfo {pages} {8938--8944} (\bibinfo {year} {1994})}\BibitemShut {NoStop}%
\bibitem [{\citenamefont {Nooijen}\ and\ \citenamefont
  {Bartlett}(1997)}]{Nooijen97}%
  \BibitemOpen
  \bibfield  {author} {\bibinfo {author} {\bibfnamefont {M.}~\bibnamefont
  {Nooijen}}\ and\ \bibinfo {author} {\bibfnamefont {R.~J.}\ \bibnamefont
  {Bartlett}},\ }\bibfield  {title} {\enquote {\bibinfo {title} {A new method
  for excited states: Similarity transformed equation-of-motion coupled-cluster
  theory},}\ }\href {\doibase 10.1063/1.474000} {\bibfield  {journal} {\bibinfo
   {journal} {J. Chem. Phys.}\ }\textbf {\bibinfo {volume} {106}},\ \bibinfo
  {pages} {6441–6448} (\bibinfo {year} {1997})}\BibitemShut {NoStop}%
\bibitem [{\citenamefont {Sattelmeyer}, \citenamefont {Schaefer~III},\ and\
  \citenamefont {Stanton}(2003)}]{Sattelmeyer03}%
  \BibitemOpen
  \bibfield  {author} {\bibinfo {author} {\bibfnamefont {K.~W.}\ \bibnamefont
  {Sattelmeyer}}, \bibinfo {author} {\bibfnamefont {H.~F.}\ \bibnamefont
  {Schaefer~III}}, \ and\ \bibinfo {author} {\bibfnamefont {J.~F.}\
  \bibnamefont {Stanton}},\ }\bibfield  {title} {\enquote {\bibinfo {title}
  {Use of 2h and 3h-p-like coupled-cluster {T}amm–{D}ancoff approaches for
  the equilibrium properties of ozone},}\ }\href {\doibase
  10.1016/S0009-2614(03)01181-3} {\bibfield  {journal} {\bibinfo  {journal}
  {Chem. Phys. Lett.}\ }\textbf {\bibinfo {volume} {378}},\ \bibinfo {pages}
  {42–46} (\bibinfo {year} {2003})}\BibitemShut {NoStop}%
\bibitem [{\citenamefont {McCurdy}\ \emph {et~al.}(1980)\citenamefont
  {McCurdy}, \citenamefont {Rescigno}, \citenamefont {Davidson},\ and\
  \citenamefont {Lauderdale}}]{McCurdy80}%
  \BibitemOpen
  \bibfield  {author} {\bibinfo {author} {\bibfnamefont {C.~W.}\ \bibnamefont
  {McCurdy}}, \bibinfo {author} {\bibfnamefont {T.~N.}\ \bibnamefont
  {Rescigno}}, \bibinfo {author} {\bibfnamefont {E.~R.}\ \bibnamefont
  {Davidson}}, \ and\ \bibinfo {author} {\bibfnamefont {J.~G.}\ \bibnamefont
  {Lauderdale}},\ }\bibfield  {title} {\enquote {\bibinfo {title}
  {Applicability of self‐consistent field techniques based on the complex
  coordinate method to metastable electronic states},}\ }\href {\doibase
  10.1063/1.440522} {\bibfield  {journal} {\bibinfo  {journal} {J. Chem.
  Phys.}\ }\textbf {\bibinfo {volume} {73}},\ \bibinfo {pages} {3268–3273}
  (\bibinfo {year} {1980})}\BibitemShut {NoStop}%
\bibitem [{\citenamefont {Rescigno}, \citenamefont {Orel},\ and\ \citenamefont
  {McCurdy}(1980)}]{Rescigno80}%
  \BibitemOpen
  \bibfield  {author} {\bibinfo {author} {\bibfnamefont {T.~N.}\ \bibnamefont
  {Rescigno}}, \bibinfo {author} {\bibfnamefont {A.~E.}\ \bibnamefont {Orel}},
  \ and\ \bibinfo {author} {\bibfnamefont {C.~W.}\ \bibnamefont {McCurdy}},\
  }\bibfield  {title} {\enquote {\bibinfo {title} {Application of complex
  coordinate scf techniques to a molecular shape resonance: The $^2\pi_g$ state
  of {N$_2^-$}},}\ }\href {\doibase 10.1063/1.440100} {\bibfield  {journal}
  {\bibinfo  {journal} {J. Chem. Phys.}\ }\textbf {\bibinfo {volume} {73}},\
  \bibinfo {pages} {6347–6348} (\bibinfo {year} {1980})}\BibitemShut
  {NoStop}%
\bibitem [{\citenamefont {Honigmann}, \citenamefont {Buenker},\ and\
  \citenamefont {Liebermann}(2006)}]{Honigmann06}%
  \BibitemOpen
  \bibfield  {author} {\bibinfo {author} {\bibfnamefont {M.}~\bibnamefont
  {Honigmann}}, \bibinfo {author} {\bibfnamefont {R.~J.}\ \bibnamefont
  {Buenker}}, \ and\ \bibinfo {author} {\bibfnamefont {H.-P.}\ \bibnamefont
  {Liebermann}},\ }\bibfield  {title} {\enquote {\bibinfo {title} {Complex
  self-consistent field and multireference single- and double-excitation
  configuration interaction calculations for the {$^2\Pi_g$} resonance state of
  {N}$_2^-$},}\ }\href {\doibase 10.1063/1.2403856} {\bibfield  {journal}
  {\bibinfo  {journal} {J. Chem. Phys.}\ }\textbf {\bibinfo {volume} {125}},\
  \bibinfo {pages} {234304} (\bibinfo {year} {2006})}\BibitemShut {NoStop}%
\bibitem [{\citenamefont {Honigmann}, \citenamefont {Liebermann},\ and\
  \citenamefont {Buenker}(2010)}]{Honigmann10}%
  \BibitemOpen
  \bibfield  {author} {\bibinfo {author} {\bibfnamefont {M.}~\bibnamefont
  {Honigmann}}, \bibinfo {author} {\bibfnamefont {H.-P.}\ \bibnamefont
  {Liebermann}}, \ and\ \bibinfo {author} {\bibfnamefont {R.~J.}\ \bibnamefont
  {Buenker}},\ }\bibfield  {title} {\enquote {\bibinfo {title} {Use of complex
  configuration interaction calculations and the stationary principle for the
  description of metastable electronic states of {HCl$^-$}},}\ }\href {\doibase
  10.1063/1.3467885} {\bibfield  {journal} {\bibinfo  {journal} {J. Chem.
  Phys.}\ }\textbf {\bibinfo {volume} {133}},\ \bibinfo {pages} {044305}
  (\bibinfo {year} {2010})}\BibitemShut {NoStop}%
\bibitem [{\citenamefont {White}, \citenamefont {McCurdy},\ and\ \citenamefont
  {Head-Gordon}(2015)}]{White15}%
  \BibitemOpen
  \bibfield  {author} {\bibinfo {author} {\bibfnamefont {A.~F.}\ \bibnamefont
  {White}}, \bibinfo {author} {\bibfnamefont {C.~W.}\ \bibnamefont {McCurdy}},
  \ and\ \bibinfo {author} {\bibfnamefont {M.}~\bibnamefont {Head-Gordon}},\
  }\bibfield  {title} {\enquote {\bibinfo {title} {Restricted and unrestricted
  non-{Hermitian} {Hartree}-{Fock}: Theory, practical considerations, and
  applications to metastable molecular anions},}\ }\href {\doibase
  10.1063/1.4928529} {\bibfield  {journal} {\bibinfo  {journal} {J. Chem.
  Phys.}\ }\textbf {\bibinfo {volume} {143}},\ \bibinfo {pages} {074103}
  (\bibinfo {year} {2015})}\BibitemShut {NoStop}%
\bibitem [{\citenamefont {White}, \citenamefont {Head-Gordon},\ and\
  \citenamefont {McCurdy}(2015)}]{White15-2}%
  \BibitemOpen
  \bibfield  {author} {\bibinfo {author} {\bibfnamefont {A.~F.}\ \bibnamefont
  {White}}, \bibinfo {author} {\bibfnamefont {M.}~\bibnamefont {Head-Gordon}},
  \ and\ \bibinfo {author} {\bibfnamefont {C.~W.}\ \bibnamefont {McCurdy}},\
  }\bibfield  {title} {\enquote {\bibinfo {title} {Complex basis functions
  revisited: Implementation with applications to carbon tetrafluoride and
  aromatic {N}-containing heterocycles within the static-exchange
  approximation},}\ }\href {\doibase 10.1063/1.4906940} {\bibfield  {journal}
  {\bibinfo  {journal} {J. Chem. Phys.}\ }\textbf {\bibinfo {volume} {142}},\
  \bibinfo {pages} {054103} (\bibinfo {year} {2015})}\BibitemShut {NoStop}%
\bibitem [{\citenamefont {White}\ \emph {et~al.}(2017)\citenamefont {White},
  \citenamefont {Epifanovsky}, \citenamefont {McCurdy},\ and\ \citenamefont
  {Head-Gordon}}]{White17}%
  \BibitemOpen
  \bibfield  {author} {\bibinfo {author} {\bibfnamefont {A.~F.}\ \bibnamefont
  {White}}, \bibinfo {author} {\bibfnamefont {E.}~\bibnamefont {Epifanovsky}},
  \bibinfo {author} {\bibfnamefont {C.~W.}\ \bibnamefont {McCurdy}}, \ and\
  \bibinfo {author} {\bibfnamefont {M.}~\bibnamefont {Head-Gordon}},\
  }\bibfield  {title} {\enquote {\bibinfo {title} {Second order
  {M}øller-{P}lesset and coupled cluster singles and doubles methods with
  complex basis functions for resonances in electron-molecule scattering},}\
  }\href {\doibase 10.1063/1.4986950} {\bibfield  {journal} {\bibinfo
  {journal} {J. Chem. Phys.}\ }\textbf {\bibinfo {volume} {146}},\ \bibinfo
  {pages} {234107} (\bibinfo {year} {2017})}\BibitemShut {NoStop}%
\bibitem [{\citenamefont {Jagau}(2018)}]{Jagau18}%
  \BibitemOpen
  \bibfield  {author} {\bibinfo {author} {\bibfnamefont {T.-C.}\ \bibnamefont
  {Jagau}},\ }\bibfield  {title} {\enquote {\bibinfo {title} {Coupled-cluster
  treatment of molecular strong-field ionization},}\ }\href {\doibase
  10.1063/1.5028179} {\bibfield  {journal} {\bibinfo  {journal} {J. Chem.
  Phys.}\ }\textbf {\bibinfo {volume} {148}},\ \bibinfo {pages} {204102}
  (\bibinfo {year} {2018})}\BibitemShut {NoStop}%
\bibitem [{\citenamefont {Hern\'andez~Vera}\ and\ \citenamefont
  {Jagau}(2019)}]{Hernandez19}%
  \BibitemOpen
  \bibfield  {author} {\bibinfo {author} {\bibfnamefont {M.}~\bibnamefont
  {Hern\'andez~Vera}}\ and\ \bibinfo {author} {\bibfnamefont {T.-C.}\
  \bibnamefont {Jagau}},\ }\bibfield  {title} {\enquote {\bibinfo {title}
  {Resolution-of-the-identity approximation for complex-scaled basis
  functions},}\ }\href {\doibase 10.1063/1.5119695} {\bibfield  {journal}
  {\bibinfo  {journal} {J. Chem. Phys.}\ }\textbf {\bibinfo {volume} {151}},\
  \bibinfo {pages} {111101} (\bibinfo {year} {2019})}\BibitemShut {NoStop}%
\bibitem [{\citenamefont {Hernández~Vera}\ and\ \citenamefont
  {Jagau}(2020)}]{Hernandez20}%
  \BibitemOpen
  \bibfield  {author} {\bibinfo {author} {\bibfnamefont {M.}~\bibnamefont
  {Hernández~Vera}}\ and\ \bibinfo {author} {\bibfnamefont {T.-C.}\
  \bibnamefont {Jagau}},\ }\bibfield  {title} {\enquote {\bibinfo {title}
  {Resolution-of-the-identity second-order {M}øller–{P}lesset perturbation
  theory with complex basis functions: Benchmark calculations and applications
  to strong-field ionization of polyacenes},}\ }\href {\doibase
  10.1063/5.0004843} {\bibfield  {journal} {\bibinfo  {journal} {J. Chem.
  Phys.}\ }\textbf {\bibinfo {volume} {152}},\ \bibinfo {pages} {174103}
  (\bibinfo {year} {2020})}\BibitemShut {NoStop}%
\bibitem [{\citenamefont {Thompson}, \citenamefont {Ochsenfeld},\ and\
  \citenamefont {Jagau}(2019)}]{Thompson19}%
  \BibitemOpen
  \bibfield  {author} {\bibinfo {author} {\bibfnamefont {T.~H.}\ \bibnamefont
  {Thompson}}, \bibinfo {author} {\bibfnamefont {C.}~\bibnamefont
  {Ochsenfeld}}, \ and\ \bibinfo {author} {\bibfnamefont {T.-C.}\ \bibnamefont
  {Jagau}},\ }\bibfield  {title} {\enquote {\bibinfo {title} {A schwarz
  inequality for complex basis function methods in non-hermitian quantum
  chemistry},}\ }\href {\doibase 10.1063/1.5123541} {\bibfield  {journal}
  {\bibinfo  {journal} {J. Chem. Phys.}\ }\textbf {\bibinfo {volume} {151}},\
  \bibinfo {pages} {184104} (\bibinfo {year} {2019})}\BibitemShut {NoStop}%
\bibitem [{\citenamefont {Matz}\ and\ \citenamefont
  {Jagau}(2022{\natexlab{a}})}]{Matz22}%
  \BibitemOpen
  \bibfield  {author} {\bibinfo {author} {\bibfnamefont {F.}~\bibnamefont
  {Matz}}\ and\ \bibinfo {author} {\bibfnamefont {T.-C.}\ \bibnamefont
  {Jagau}},\ }\bibfield  {title} {\enquote {\bibinfo {title} {Molecular {Auger}
  decay rates from complex-variable coupled-cluster theory},}\ }\href {\doibase
  10.1063/5.0075646} {\bibfield  {journal} {\bibinfo  {journal} {J. Chem.
  Phys.}\ }\textbf {\bibinfo {volume} {156}},\ \bibinfo {pages} {114117}
  (\bibinfo {year} {2022}{\natexlab{a}})}\BibitemShut {NoStop}%
\bibitem [{\citenamefont {Matz}\ and\ \citenamefont
  {Jagau}(2022{\natexlab{b}})}]{Matz22-2}%
  \BibitemOpen
  \bibfield  {author} {\bibinfo {author} {\bibfnamefont {F.}~\bibnamefont
  {Matz}}\ and\ \bibinfo {author} {\bibfnamefont {T.-C.}\ \bibnamefont
  {Jagau}},\ }\bibfield  {title} {\enquote {\bibinfo {title} {Channel-specific
  core-valence projectors for determining partial {Auger} decay widths},}\
  }\href {\doibase 10.1080/00268976.2022.2105270} {\bibfield  {journal}
  {\bibinfo  {journal} {Mol. Phys.}\ }\textbf {\bibinfo {volume} {120}},\
  \bibinfo {pages} {e2105270} (\bibinfo {year}
  {2022}{\natexlab{b}})}\BibitemShut {NoStop}%
\bibitem [{\citenamefont {Jayadev}\ \emph {et~al.}(2023)\citenamefont
  {Jayadev}, \citenamefont {Ferino-P\'erez}, \citenamefont {Matz},
  \citenamefont {Krylov},\ and\ \citenamefont {Jagau}}]{Jayadev23}%
  \BibitemOpen
  \bibfield  {author} {\bibinfo {author} {\bibfnamefont {N.~K.}\ \bibnamefont
  {Jayadev}}, \bibinfo {author} {\bibfnamefont {A.}~\bibnamefont
  {Ferino-P\'erez}}, \bibinfo {author} {\bibfnamefont {F.}~\bibnamefont
  {Matz}}, \bibinfo {author} {\bibfnamefont {A.~I.}\ \bibnamefont {Krylov}}, \
  and\ \bibinfo {author} {\bibfnamefont {T.-C.}\ \bibnamefont {Jagau}},\
  }\bibfield  {title} {\enquote {\bibinfo {title} {The {Auger} spectrum of
  benzene},}\ }\href {\doibase 10.1063/5.0138674} {\bibfield  {journal}
  {\bibinfo  {journal} {J. Chem. Phys.}\ }\textbf {\bibinfo {volume} {158}},\
  \bibinfo {pages} {064109} (\bibinfo {year} {2023})}\BibitemShut {NoStop}%
\bibitem [{\citenamefont {Skomorowski}\ and\ \citenamefont
  {Krylov}(2021{\natexlab{b}})}]{Skomorowski21-2}%
  \BibitemOpen
  \bibfield  {author} {\bibinfo {author} {\bibfnamefont {W.}~\bibnamefont
  {Skomorowski}}\ and\ \bibinfo {author} {\bibfnamefont {A.~I.}\ \bibnamefont
  {Krylov}},\ }\bibfield  {title} {\enquote {\bibinfo {title} {Feshbach-{Fano}
  approach for calculation of {Auger} decay rates using equation-of-motion
  coupled-cluster wave functions. {II}. {Numerical} examples and benchmarks},}\
  }\href {\doibase 10.1063/5.0036977} {\bibfield  {journal} {\bibinfo
  {journal} {J. Chem. Phys.}\ }\textbf {\bibinfo {volume} {154}},\ \bibinfo
  {pages} {084125} (\bibinfo {year} {2021}{\natexlab{b}})}\BibitemShut
  {NoStop}%
\bibitem [{\citenamefont {Aguilar}\ and\ \citenamefont
  {Combes}(1971)}]{Aguilar71}%
  \BibitemOpen
  \bibfield  {author} {\bibinfo {author} {\bibfnamefont {J.}~\bibnamefont
  {Aguilar}}\ and\ \bibinfo {author} {\bibfnamefont {J.~M.}\ \bibnamefont
  {Combes}},\ }\bibfield  {title} {\enquote {\bibinfo {title} {A class of
  analytic perturbations for one-body {S}chrödinger {H}amiltonians},}\ }\href
  {\doibase 10.1007/BF01877510} {\bibfield  {journal} {\bibinfo  {journal}
  {Commun. Math. Phys.}\ }\textbf {\bibinfo {volume} {22}},\ \bibinfo {pages}
  {269–279} (\bibinfo {year} {1971})}\BibitemShut {NoStop}%
\bibitem [{\citenamefont {Balslev}\ and\ \citenamefont
  {Combes}(1971)}]{Balslev71}%
  \BibitemOpen
  \bibfield  {author} {\bibinfo {author} {\bibfnamefont {E.}~\bibnamefont
  {Balslev}}\ and\ \bibinfo {author} {\bibfnamefont {J.~M.}\ \bibnamefont
  {Combes}},\ }\bibfield  {title} {\enquote {\bibinfo {title} {Spectral
  properties of many-body {Schrödinger} operators with dilatation-analytic
  interactions},}\ }\href {\doibase 10.1007/BF01877511} {\bibfield  {journal}
  {\bibinfo  {journal} {Commun. Math. Phys.}\ }\textbf {\bibinfo {volume}
  {22}},\ \bibinfo {pages} {280–294} (\bibinfo {year} {1971})}\BibitemShut
  {NoStop}%
\bibitem [{\citenamefont {Gilbert}, \citenamefont {Besley},\ and\ \citenamefont
  {Gill}(2008)}]{Gilbert08}%
  \BibitemOpen
  \bibfield  {author} {\bibinfo {author} {\bibfnamefont {A.~T.~B.}\
  \bibnamefont {Gilbert}}, \bibinfo {author} {\bibfnamefont {N.~A.}\
  \bibnamefont {Besley}}, \ and\ \bibinfo {author} {\bibfnamefont {P.~M.~W.}\
  \bibnamefont {Gill}},\ }\bibfield  {title} {\enquote {\bibinfo {title}
  {Self-consistent field calculations of excited states using the maximum
  overlap method {(MOM)}},}\ }\href {\doibase 10.1021/jp801738f} {\bibfield
  {journal} {\bibinfo  {journal} {J. Phys. Chem. A}\ }\textbf {\bibinfo
  {volume} {112}},\ \bibinfo {pages} {13164–13171} (\bibinfo {year}
  {2008})}\BibitemShut {NoStop}%
\bibitem [{\citenamefont {Moiseyev}, \citenamefont {Certain},\ and\
  \citenamefont {Weinhold}(1978)}]{Moiseyev78}%
  \BibitemOpen
  \bibfield  {author} {\bibinfo {author} {\bibfnamefont {N.}~\bibnamefont
  {Moiseyev}}, \bibinfo {author} {\bibfnamefont {P.}~\bibnamefont {Certain}}, \
  and\ \bibinfo {author} {\bibfnamefont {F.}~\bibnamefont {Weinhold}},\
  }\bibfield  {title} {\enquote {\bibinfo {title} {Resonance properties of
  complex-rotated hamiltonians},}\ }\href {\doibase 10.1080/00268977800102631}
  {\bibfield  {journal} {\bibinfo  {journal} {Mol. Phys.}\ }\textbf {\bibinfo
  {volume} {36}},\ \bibinfo {pages} {1613–1630} (\bibinfo {year}
  {1978})}\BibitemShut {NoStop}%
\bibitem [{\citenamefont {Epifanovsky}\ \emph {et~al.}(2021)\citenamefont
  {Epifanovsky}, \citenamefont {Gilbert}, \citenamefont {Feng}, \citenamefont
  {Lee}, \citenamefont {Mao}, \citenamefont {Mardirossian}, \citenamefont
  {Pokhilko}, \citenamefont {White}, \citenamefont {Coons}, \citenamefont
  {Dempwolff}, \citenamefont {Gan}, \citenamefont {Hait}, \citenamefont {Horn},
  \citenamefont {Jacobson}, \citenamefont {Kaliman}, \citenamefont {Kussmann},
  \citenamefont {Lange}, \citenamefont {Lao}, \citenamefont {Levine},
  \citenamefont {Liu}, \citenamefont {McKenzie}, \citenamefont {Morrison},
  \citenamefont {Nanda}, \citenamefont {Plasser}, \citenamefont {Rehn},
  \citenamefont {Vidal}, \citenamefont {You}, \citenamefont {Zhu},
  \citenamefont {Alam}, \citenamefont {Albrecht}, \citenamefont {Aldossary},
  \citenamefont {Alguire}, \citenamefont {Andersen}, \citenamefont {Athavale},
  \citenamefont {Barton}, \citenamefont {Begam}, \citenamefont {Behn},
  \citenamefont {Bellonzi}, \citenamefont {Bernard}, \citenamefont {Berquist},
  \citenamefont {Burton}, \citenamefont {Carreras}, \citenamefont
  {Carter-Fenk}, \citenamefont {Chakraborty}, \citenamefont {Chien},
  \citenamefont {Closser}, \citenamefont {Cofer-Shabica}, \citenamefont
  {Dasgupta}, \citenamefont {de~Wergifosse}, \citenamefont {Deng},
  \citenamefont {Diedenhofen}, \citenamefont {Do}, \citenamefont {Ehlert},
  \citenamefont {Fang}, \citenamefont {Fatehi}, \citenamefont {Feng},
  \citenamefont {Friedhoff}, \citenamefont {Gayvert}, \citenamefont {Ge},
  \citenamefont {Gidofalvi}, \citenamefont {Goldey}, \citenamefont {Gomes},
  \citenamefont {González-Espinoza}, \citenamefont {Gulania}, \citenamefont
  {Gunina}, \citenamefont {Hanson-Heine}, \citenamefont {Harbach},
  \citenamefont {Hauser}, \citenamefont {Herbst}, \citenamefont
  {Hernández~Vera}, \citenamefont {Hodecker}, \citenamefont {Holden},
  \citenamefont {Houck}, \citenamefont {Huang}, \citenamefont {Hui},
  \citenamefont {Huynh}, \citenamefont {Ivanov}, \citenamefont {Jász},
  \citenamefont {Ji}, \citenamefont {Jiang}, \citenamefont {Kaduk},
  \citenamefont {Kähler}, \citenamefont {Khistyaev}, \citenamefont {Kim},
  \citenamefont {Kis}, \citenamefont {Klunzinger}, \citenamefont
  {Koczor-Benda}, \citenamefont {Koh}, \citenamefont {Kosenkov}, \citenamefont
  {Koulias}, \citenamefont {Kowalczyk}, \citenamefont {Krauter}, \citenamefont
  {Kue}, \citenamefont {Kunitsa}, \citenamefont {Kus}, \citenamefont
  {Ladjánszki}, \citenamefont {Landau}, \citenamefont {Lawler}, \citenamefont
  {Lefrancois}, \citenamefont {Lehtola}, \citenamefont {Li}, \citenamefont
  {Li}, \citenamefont {Liang}, \citenamefont {Liebenthal}, \citenamefont {Lin},
  \citenamefont {Lin}, \citenamefont {Liu}, \citenamefont {Liu}, \citenamefont
  {Loipersberger}, \citenamefont {Luenser}, \citenamefont {Manjanath},
  \citenamefont {Manohar}, \citenamefont {Mansoor}, \citenamefont {Manzer},
  \citenamefont {Mao}, \citenamefont {Marenich}, \citenamefont {Markovich},
  \citenamefont {Mason}, \citenamefont {Maurer}, \citenamefont {McLaughlin},
  \citenamefont {Menger}, \citenamefont {Mewes}, \citenamefont {Mewes},
  \citenamefont {Morgante}, \citenamefont {Mullinax}, \citenamefont
  {Oosterbaan}, \citenamefont {Paran}, \citenamefont {Paul}, \citenamefont
  {Paul}, \citenamefont {Pavošević}, \citenamefont {Pei}, \citenamefont
  {Prager}, \citenamefont {Proynov}, \citenamefont {Rák}, \citenamefont
  {Ramos-Cordoba}, \citenamefont {Rana}, \citenamefont {Rask}, \citenamefont
  {Rettig}, \citenamefont {Richard}, \citenamefont {Rob}, \citenamefont
  {Rossomme}, \citenamefont {Scheele}, \citenamefont {Scheurer}, \citenamefont
  {Schneider}, \citenamefont {Sergueev}, \citenamefont {Sharada}, \citenamefont
  {Skomorowski}, \citenamefont {Small}, \citenamefont {Stein}, \citenamefont
  {Su}, \citenamefont {Sundstrom}, \citenamefont {Tao}, \citenamefont
  {Thirman}, \citenamefont {Tornai}, \citenamefont {Tsuchimochi}, \citenamefont
  {Tubman}, \citenamefont {Veccham}, \citenamefont {Vydrov}, \citenamefont
  {Wenzel}, \citenamefont {Witte}, \citenamefont {Yamada}, \citenamefont {Yao},
  \citenamefont {Yeganeh}, \citenamefont {Yost}, \citenamefont {Zech},
  \citenamefont {Zhang}, \citenamefont {Zhang}, \citenamefont {Zhang},
  \citenamefont {Zuev}, \citenamefont {Aspuru-Guzik}, \citenamefont {Bell},
  \citenamefont {Besley}, \citenamefont {Bravaya}, \citenamefont {Brooks},
  \citenamefont {Casanova}, \citenamefont {Chai}, \citenamefont {Coriani},
  \citenamefont {Cramer}, \citenamefont {Cserey}, \citenamefont {DePrince},
  \citenamefont {DiStasio}, \citenamefont {Dreuw}, \citenamefont {Dunietz},
  \citenamefont {Furlani}, \citenamefont {Goddard}, \citenamefont
  {Hammes-Schiffer}, \citenamefont {Head-Gordon}, \citenamefont {Hehre},
  \citenamefont {Hsu}, \citenamefont {Jagau}, \citenamefont {Jung},
  \citenamefont {Klamt}, \citenamefont {Kong}, \citenamefont {Lambrecht},
  \citenamefont {Liang}, \citenamefont {Mayhall}, \citenamefont {McCurdy},
  \citenamefont {Neaton}, \citenamefont {Ochsenfeld}, \citenamefont {Parkhill},
  \citenamefont {Peverati}, \citenamefont {Rassolov}, \citenamefont {Shao},
  \citenamefont {Slipchenko}, \citenamefont {Stauch}, \citenamefont {Steele},
  \citenamefont {Subotnik}, \citenamefont {Thom}, \citenamefont {Tkatchenko},
  \citenamefont {Truhlar}, \citenamefont {Van~Voorhis}, \citenamefont
  {Wesolowski}, \citenamefont {Whaley}, \citenamefont {Woodcock}, \citenamefont
  {Zimmerman}, \citenamefont {Faraji}, \citenamefont {Gill}, \citenamefont
  {Head-Gordon}, \citenamefont {Herbert},\ and\ \citenamefont
  {Krylov}}]{qchem5}%
  \BibitemOpen
  \bibfield  {author} {\bibinfo {author} {\bibfnamefont {E.}~\bibnamefont
  {Epifanovsky}}, \bibinfo {author} {\bibfnamefont {A.~T.~B.}\ \bibnamefont
  {Gilbert}}, \bibinfo {author} {\bibfnamefont {X.}~\bibnamefont {Feng}},
  \bibinfo {author} {\bibfnamefont {J.}~\bibnamefont {Lee}}, \bibinfo {author}
  {\bibfnamefont {Y.}~\bibnamefont {Mao}}, \bibinfo {author} {\bibfnamefont
  {N.}~\bibnamefont {Mardirossian}}, \bibinfo {author} {\bibfnamefont
  {P.}~\bibnamefont {Pokhilko}}, \bibinfo {author} {\bibfnamefont {A.~F.}\
  \bibnamefont {White}}, \bibinfo {author} {\bibfnamefont {M.~P.}\ \bibnamefont
  {Coons}}, \bibinfo {author} {\bibfnamefont {A.~L.}\ \bibnamefont
  {Dempwolff}}, \bibinfo {author} {\bibfnamefont {Z.}~\bibnamefont {Gan}},
  \bibinfo {author} {\bibfnamefont {D.}~\bibnamefont {Hait}}, \bibinfo {author}
  {\bibfnamefont {P.~R.}\ \bibnamefont {Horn}}, \bibinfo {author}
  {\bibfnamefont {L.~D.}\ \bibnamefont {Jacobson}}, \bibinfo {author}
  {\bibfnamefont {I.}~\bibnamefont {Kaliman}}, \bibinfo {author} {\bibfnamefont
  {J.}~\bibnamefont {Kussmann}}, \bibinfo {author} {\bibfnamefont {A.~W.}\
  \bibnamefont {Lange}}, \bibinfo {author} {\bibfnamefont {K.~U.}\ \bibnamefont
  {Lao}}, \bibinfo {author} {\bibfnamefont {D.~S.}\ \bibnamefont {Levine}},
  \bibinfo {author} {\bibfnamefont {J.}~\bibnamefont {Liu}}, \bibinfo {author}
  {\bibfnamefont {S.~C.}\ \bibnamefont {McKenzie}}, \bibinfo {author}
  {\bibfnamefont {A.~F.}\ \bibnamefont {Morrison}}, \bibinfo {author}
  {\bibfnamefont {K.~D.}\ \bibnamefont {Nanda}}, \bibinfo {author}
  {\bibfnamefont {F.}~\bibnamefont {Plasser}}, \bibinfo {author} {\bibfnamefont
  {D.~R.}\ \bibnamefont {Rehn}}, \bibinfo {author} {\bibfnamefont {M.~L.}\
  \bibnamefont {Vidal}}, \bibinfo {author} {\bibfnamefont {Z.-Q.}\ \bibnamefont
  {You}}, \bibinfo {author} {\bibfnamefont {Y.}~\bibnamefont {Zhu}}, \bibinfo
  {author} {\bibfnamefont {B.}~\bibnamefont {Alam}}, \bibinfo {author}
  {\bibfnamefont {B.~J.}\ \bibnamefont {Albrecht}}, \bibinfo {author}
  {\bibfnamefont {A.}~\bibnamefont {Aldossary}}, \bibinfo {author}
  {\bibfnamefont {E.}~\bibnamefont {Alguire}}, \bibinfo {author} {\bibfnamefont
  {J.~H.}\ \bibnamefont {Andersen}}, \bibinfo {author} {\bibfnamefont
  {V.}~\bibnamefont {Athavale}}, \bibinfo {author} {\bibfnamefont
  {D.}~\bibnamefont {Barton}}, \bibinfo {author} {\bibfnamefont
  {K.}~\bibnamefont {Begam}}, \bibinfo {author} {\bibfnamefont
  {A.}~\bibnamefont {Behn}}, \bibinfo {author} {\bibfnamefont {N.}~\bibnamefont
  {Bellonzi}}, \bibinfo {author} {\bibfnamefont {Y.~A.}\ \bibnamefont
  {Bernard}}, \bibinfo {author} {\bibfnamefont {E.~J.}\ \bibnamefont
  {Berquist}}, \bibinfo {author} {\bibfnamefont {H.~G.~A.}\ \bibnamefont
  {Burton}}, \bibinfo {author} {\bibfnamefont {A.}~\bibnamefont {Carreras}},
  \bibinfo {author} {\bibfnamefont {K.}~\bibnamefont {Carter-Fenk}}, \bibinfo
  {author} {\bibfnamefont {R.}~\bibnamefont {Chakraborty}}, \bibinfo {author}
  {\bibfnamefont {A.~D.}\ \bibnamefont {Chien}}, \bibinfo {author}
  {\bibfnamefont {K.~D.}\ \bibnamefont {Closser}}, \bibinfo {author}
  {\bibfnamefont {V.}~\bibnamefont {Cofer-Shabica}}, \bibinfo {author}
  {\bibfnamefont {S.}~\bibnamefont {Dasgupta}}, \bibinfo {author}
  {\bibfnamefont {M.}~\bibnamefont {de~Wergifosse}}, \bibinfo {author}
  {\bibfnamefont {J.}~\bibnamefont {Deng}}, \bibinfo {author} {\bibfnamefont
  {M.}~\bibnamefont {Diedenhofen}}, \bibinfo {author} {\bibfnamefont
  {H.}~\bibnamefont {Do}}, \bibinfo {author} {\bibfnamefont {S.}~\bibnamefont
  {Ehlert}}, \bibinfo {author} {\bibfnamefont {P.-T.}\ \bibnamefont {Fang}},
  \bibinfo {author} {\bibfnamefont {S.}~\bibnamefont {Fatehi}}, \bibinfo
  {author} {\bibfnamefont {Q.}~\bibnamefont {Feng}}, \bibinfo {author}
  {\bibfnamefont {T.}~\bibnamefont {Friedhoff}}, \bibinfo {author}
  {\bibfnamefont {J.}~\bibnamefont {Gayvert}}, \bibinfo {author} {\bibfnamefont
  {Q.}~\bibnamefont {Ge}}, \bibinfo {author} {\bibfnamefont {G.}~\bibnamefont
  {Gidofalvi}}, \bibinfo {author} {\bibfnamefont {M.}~\bibnamefont {Goldey}},
  \bibinfo {author} {\bibfnamefont {J.}~\bibnamefont {Gomes}}, \bibinfo
  {author} {\bibfnamefont {C.~E.}\ \bibnamefont {González-Espinoza}}, \bibinfo
  {author} {\bibfnamefont {S.}~\bibnamefont {Gulania}}, \bibinfo {author}
  {\bibfnamefont {A.~O.}\ \bibnamefont {Gunina}}, \bibinfo {author}
  {\bibfnamefont {M.~W.~D.}\ \bibnamefont {Hanson-Heine}}, \bibinfo {author}
  {\bibfnamefont {P.~H.~P.}\ \bibnamefont {Harbach}}, \bibinfo {author}
  {\bibfnamefont {A.}~\bibnamefont {Hauser}}, \bibinfo {author} {\bibfnamefont
  {M.~F.}\ \bibnamefont {Herbst}}, \bibinfo {author} {\bibfnamefont
  {M.}~\bibnamefont {Hernández~Vera}}, \bibinfo {author} {\bibfnamefont
  {M.}~\bibnamefont {Hodecker}}, \bibinfo {author} {\bibfnamefont {Z.~C.}\
  \bibnamefont {Holden}}, \bibinfo {author} {\bibfnamefont {S.}~\bibnamefont
  {Houck}}, \bibinfo {author} {\bibfnamefont {X.}~\bibnamefont {Huang}},
  \bibinfo {author} {\bibfnamefont {K.}~\bibnamefont {Hui}}, \bibinfo {author}
  {\bibfnamefont {B.~C.}\ \bibnamefont {Huynh}}, \bibinfo {author}
  {\bibfnamefont {M.}~\bibnamefont {Ivanov}}, \bibinfo {author} {\bibfnamefont
  {Á.}~\bibnamefont {Jász}}, \bibinfo {author} {\bibfnamefont
  {H.}~\bibnamefont {Ji}}, \bibinfo {author} {\bibfnamefont {H.}~\bibnamefont
  {Jiang}}, \bibinfo {author} {\bibfnamefont {B.}~\bibnamefont {Kaduk}},
  \bibinfo {author} {\bibfnamefont {S.}~\bibnamefont {Kähler}}, \bibinfo
  {author} {\bibfnamefont {K.}~\bibnamefont {Khistyaev}}, \bibinfo {author}
  {\bibfnamefont {J.}~\bibnamefont {Kim}}, \bibinfo {author} {\bibfnamefont
  {G.}~\bibnamefont {Kis}}, \bibinfo {author} {\bibfnamefont {P.}~\bibnamefont
  {Klunzinger}}, \bibinfo {author} {\bibfnamefont {Z.}~\bibnamefont
  {Koczor-Benda}}, \bibinfo {author} {\bibfnamefont {J.~H.}\ \bibnamefont
  {Koh}}, \bibinfo {author} {\bibfnamefont {D.}~\bibnamefont {Kosenkov}},
  \bibinfo {author} {\bibfnamefont {L.}~\bibnamefont {Koulias}}, \bibinfo
  {author} {\bibfnamefont {T.}~\bibnamefont {Kowalczyk}}, \bibinfo {author}
  {\bibfnamefont {C.~M.}\ \bibnamefont {Krauter}}, \bibinfo {author}
  {\bibfnamefont {K.}~\bibnamefont {Kue}}, \bibinfo {author} {\bibfnamefont
  {A.}~\bibnamefont {Kunitsa}}, \bibinfo {author} {\bibfnamefont
  {T.}~\bibnamefont {Kus}}, \bibinfo {author} {\bibfnamefont {I.}~\bibnamefont
  {Ladjánszki}}, \bibinfo {author} {\bibfnamefont {A.}~\bibnamefont {Landau}},
  \bibinfo {author} {\bibfnamefont {K.~V.}\ \bibnamefont {Lawler}}, \bibinfo
  {author} {\bibfnamefont {D.}~\bibnamefont {Lefrancois}}, \bibinfo {author}
  {\bibfnamefont {S.}~\bibnamefont {Lehtola}}, \bibinfo {author} {\bibfnamefont
  {R.~R.}\ \bibnamefont {Li}}, \bibinfo {author} {\bibfnamefont {Y.-P.}\
  \bibnamefont {Li}}, \bibinfo {author} {\bibfnamefont {J.}~\bibnamefont
  {Liang}}, \bibinfo {author} {\bibfnamefont {M.}~\bibnamefont {Liebenthal}},
  \bibinfo {author} {\bibfnamefont {H.-H.}\ \bibnamefont {Lin}}, \bibinfo
  {author} {\bibfnamefont {Y.-S.}\ \bibnamefont {Lin}}, \bibinfo {author}
  {\bibfnamefont {F.}~\bibnamefont {Liu}}, \bibinfo {author} {\bibfnamefont
  {K.-Y.}\ \bibnamefont {Liu}}, \bibinfo {author} {\bibfnamefont
  {M.}~\bibnamefont {Loipersberger}}, \bibinfo {author} {\bibfnamefont
  {A.}~\bibnamefont {Luenser}}, \bibinfo {author} {\bibfnamefont
  {A.}~\bibnamefont {Manjanath}}, \bibinfo {author} {\bibfnamefont
  {P.}~\bibnamefont {Manohar}}, \bibinfo {author} {\bibfnamefont
  {E.}~\bibnamefont {Mansoor}}, \bibinfo {author} {\bibfnamefont {S.~F.}\
  \bibnamefont {Manzer}}, \bibinfo {author} {\bibfnamefont {S.-P.}\
  \bibnamefont {Mao}}, \bibinfo {author} {\bibfnamefont {A.~V.}\ \bibnamefont
  {Marenich}}, \bibinfo {author} {\bibfnamefont {T.}~\bibnamefont {Markovich}},
  \bibinfo {author} {\bibfnamefont {S.}~\bibnamefont {Mason}}, \bibinfo
  {author} {\bibfnamefont {S.~A.}\ \bibnamefont {Maurer}}, \bibinfo {author}
  {\bibfnamefont {P.~F.}\ \bibnamefont {McLaughlin}}, \bibinfo {author}
  {\bibfnamefont {M.~F. S.~J.}\ \bibnamefont {Menger}}, \bibinfo {author}
  {\bibfnamefont {J.-M.}\ \bibnamefont {Mewes}}, \bibinfo {author}
  {\bibfnamefont {S.~A.}\ \bibnamefont {Mewes}}, \bibinfo {author}
  {\bibfnamefont {P.}~\bibnamefont {Morgante}}, \bibinfo {author}
  {\bibfnamefont {J.~W.}\ \bibnamefont {Mullinax}}, \bibinfo {author}
  {\bibfnamefont {K.~J.}\ \bibnamefont {Oosterbaan}}, \bibinfo {author}
  {\bibfnamefont {G.}~\bibnamefont {Paran}}, \bibinfo {author} {\bibfnamefont
  {A.~C.}\ \bibnamefont {Paul}}, \bibinfo {author} {\bibfnamefont {S.~K.}\
  \bibnamefont {Paul}}, \bibinfo {author} {\bibfnamefont {F.}~\bibnamefont
  {Pavošević}}, \bibinfo {author} {\bibfnamefont {Z.}~\bibnamefont {Pei}},
  \bibinfo {author} {\bibfnamefont {S.}~\bibnamefont {Prager}}, \bibinfo
  {author} {\bibfnamefont {E.~I.}\ \bibnamefont {Proynov}}, \bibinfo {author}
  {\bibfnamefont {Á.}~\bibnamefont {Rák}}, \bibinfo {author} {\bibfnamefont
  {E.}~\bibnamefont {Ramos-Cordoba}}, \bibinfo {author} {\bibfnamefont
  {B.}~\bibnamefont {Rana}}, \bibinfo {author} {\bibfnamefont {A.~E.}\
  \bibnamefont {Rask}}, \bibinfo {author} {\bibfnamefont {A.}~\bibnamefont
  {Rettig}}, \bibinfo {author} {\bibfnamefont {R.~M.}\ \bibnamefont {Richard}},
  \bibinfo {author} {\bibfnamefont {F.}~\bibnamefont {Rob}}, \bibinfo {author}
  {\bibfnamefont {E.}~\bibnamefont {Rossomme}}, \bibinfo {author}
  {\bibfnamefont {T.}~\bibnamefont {Scheele}}, \bibinfo {author} {\bibfnamefont
  {M.}~\bibnamefont {Scheurer}}, \bibinfo {author} {\bibfnamefont
  {M.}~\bibnamefont {Schneider}}, \bibinfo {author} {\bibfnamefont
  {N.}~\bibnamefont {Sergueev}}, \bibinfo {author} {\bibfnamefont {S.~M.}\
  \bibnamefont {Sharada}}, \bibinfo {author} {\bibfnamefont {W.}~\bibnamefont
  {Skomorowski}}, \bibinfo {author} {\bibfnamefont {D.~W.}\ \bibnamefont
  {Small}}, \bibinfo {author} {\bibfnamefont {C.~J.}\ \bibnamefont {Stein}},
  \bibinfo {author} {\bibfnamefont {Y.-C.}\ \bibnamefont {Su}}, \bibinfo
  {author} {\bibfnamefont {E.~J.}\ \bibnamefont {Sundstrom}}, \bibinfo {author}
  {\bibfnamefont {Z.}~\bibnamefont {Tao}}, \bibinfo {author} {\bibfnamefont
  {J.}~\bibnamefont {Thirman}}, \bibinfo {author} {\bibfnamefont {G.~J.}\
  \bibnamefont {Tornai}}, \bibinfo {author} {\bibfnamefont {T.}~\bibnamefont
  {Tsuchimochi}}, \bibinfo {author} {\bibfnamefont {N.~M.}\ \bibnamefont
  {Tubman}}, \bibinfo {author} {\bibfnamefont {S.~P.}\ \bibnamefont {Veccham}},
  \bibinfo {author} {\bibfnamefont {O.}~\bibnamefont {Vydrov}}, \bibinfo
  {author} {\bibfnamefont {J.}~\bibnamefont {Wenzel}}, \bibinfo {author}
  {\bibfnamefont {J.}~\bibnamefont {Witte}}, \bibinfo {author} {\bibfnamefont
  {A.}~\bibnamefont {Yamada}}, \bibinfo {author} {\bibfnamefont
  {K.}~\bibnamefont {Yao}}, \bibinfo {author} {\bibfnamefont {S.}~\bibnamefont
  {Yeganeh}}, \bibinfo {author} {\bibfnamefont {S.~R.}\ \bibnamefont {Yost}},
  \bibinfo {author} {\bibfnamefont {A.}~\bibnamefont {Zech}}, \bibinfo {author}
  {\bibfnamefont {I.~Y.}\ \bibnamefont {Zhang}}, \bibinfo {author}
  {\bibfnamefont {X.}~\bibnamefont {Zhang}}, \bibinfo {author} {\bibfnamefont
  {Y.}~\bibnamefont {Zhang}}, \bibinfo {author} {\bibfnamefont
  {D.}~\bibnamefont {Zuev}}, \bibinfo {author} {\bibfnamefont {A.}~\bibnamefont
  {Aspuru-Guzik}}, \bibinfo {author} {\bibfnamefont {A.~T.}\ \bibnamefont
  {Bell}}, \bibinfo {author} {\bibfnamefont {N.~A.}\ \bibnamefont {Besley}},
  \bibinfo {author} {\bibfnamefont {K.~B.}\ \bibnamefont {Bravaya}}, \bibinfo
  {author} {\bibfnamefont {B.~R.}\ \bibnamefont {Brooks}}, \bibinfo {author}
  {\bibfnamefont {D.}~\bibnamefont {Casanova}}, \bibinfo {author}
  {\bibfnamefont {J.-D.}\ \bibnamefont {Chai}}, \bibinfo {author}
  {\bibfnamefont {S.}~\bibnamefont {Coriani}}, \bibinfo {author} {\bibfnamefont
  {C.~J.}\ \bibnamefont {Cramer}}, \bibinfo {author} {\bibfnamefont
  {G.}~\bibnamefont {Cserey}}, \bibinfo {author} {\bibfnamefont {A.~E.}\
  \bibnamefont {DePrince}}, \bibinfo {author} {\bibfnamefont {R.~A.}\
  \bibnamefont {DiStasio}}, \bibinfo {author} {\bibfnamefont {A.}~\bibnamefont
  {Dreuw}}, \bibinfo {author} {\bibfnamefont {B.~D.}\ \bibnamefont {Dunietz}},
  \bibinfo {author} {\bibfnamefont {T.~R.}\ \bibnamefont {Furlani}}, \bibinfo
  {author} {\bibfnamefont {W.~A.}\ \bibnamefont {Goddard}}, \bibinfo {author}
  {\bibfnamefont {S.}~\bibnamefont {Hammes-Schiffer}}, \bibinfo {author}
  {\bibfnamefont {T.}~\bibnamefont {Head-Gordon}}, \bibinfo {author}
  {\bibfnamefont {W.~J.}\ \bibnamefont {Hehre}}, \bibinfo {author}
  {\bibfnamefont {C.-P.}\ \bibnamefont {Hsu}}, \bibinfo {author} {\bibfnamefont
  {T.-C.}\ \bibnamefont {Jagau}}, \bibinfo {author} {\bibfnamefont
  {Y.}~\bibnamefont {Jung}}, \bibinfo {author} {\bibfnamefont {A.}~\bibnamefont
  {Klamt}}, \bibinfo {author} {\bibfnamefont {J.}~\bibnamefont {Kong}},
  \bibinfo {author} {\bibfnamefont {D.~S.}\ \bibnamefont {Lambrecht}}, \bibinfo
  {author} {\bibfnamefont {W.}~\bibnamefont {Liang}}, \bibinfo {author}
  {\bibfnamefont {N.~J.}\ \bibnamefont {Mayhall}}, \bibinfo {author}
  {\bibfnamefont {C.~W.}\ \bibnamefont {McCurdy}}, \bibinfo {author}
  {\bibfnamefont {J.~B.}\ \bibnamefont {Neaton}}, \bibinfo {author}
  {\bibfnamefont {C.}~\bibnamefont {Ochsenfeld}}, \bibinfo {author}
  {\bibfnamefont {J.~A.}\ \bibnamefont {Parkhill}}, \bibinfo {author}
  {\bibfnamefont {R.}~\bibnamefont {Peverati}}, \bibinfo {author}
  {\bibfnamefont {V.~A.}\ \bibnamefont {Rassolov}}, \bibinfo {author}
  {\bibfnamefont {Y.}~\bibnamefont {Shao}}, \bibinfo {author} {\bibfnamefont
  {L.~V.}\ \bibnamefont {Slipchenko}}, \bibinfo {author} {\bibfnamefont
  {T.}~\bibnamefont {Stauch}}, \bibinfo {author} {\bibfnamefont {R.~P.}\
  \bibnamefont {Steele}}, \bibinfo {author} {\bibfnamefont {J.~E.}\
  \bibnamefont {Subotnik}}, \bibinfo {author} {\bibfnamefont {A.~J.~W.}\
  \bibnamefont {Thom}}, \bibinfo {author} {\bibfnamefont {A.}~\bibnamefont
  {Tkatchenko}}, \bibinfo {author} {\bibfnamefont {D.~G.}\ \bibnamefont
  {Truhlar}}, \bibinfo {author} {\bibfnamefont {T.}~\bibnamefont
  {Van~Voorhis}}, \bibinfo {author} {\bibfnamefont {T.~A.}\ \bibnamefont
  {Wesolowski}}, \bibinfo {author} {\bibfnamefont {K.~B.}\ \bibnamefont
  {Whaley}}, \bibinfo {author} {\bibfnamefont {H.~L.}\ \bibnamefont
  {Woodcock}}, \bibinfo {author} {\bibfnamefont {P.~M.}\ \bibnamefont
  {Zimmerman}}, \bibinfo {author} {\bibfnamefont {S.}~\bibnamefont {Faraji}},
  \bibinfo {author} {\bibfnamefont {P.~M.~W.}\ \bibnamefont {Gill}}, \bibinfo
  {author} {\bibfnamefont {M.}~\bibnamefont {Head-Gordon}}, \bibinfo {author}
  {\bibfnamefont {J.~M.}\ \bibnamefont {Herbert}}, \ and\ \bibinfo {author}
  {\bibfnamefont {A.~I.}\ \bibnamefont {Krylov}},\ }\bibfield  {title}
  {\enquote {\bibinfo {title} {Software for the frontiers of quantum chemistry:
  An overview of developments in the {Q-C}hem 5 package},}\ }\href {\doibase
  10.1063/5.0055522} {\bibfield  {journal} {\bibinfo  {journal} {J. Chem.
  Phys.}\ }\textbf {\bibinfo {volume} {155}},\ \bibinfo {pages} {084801}
  (\bibinfo {year} {2021})}\BibitemShut {NoStop}%
\bibitem [{\citenamefont {Zuev}\ \emph {et~al.}(2014)\citenamefont {Zuev},
  \citenamefont {Jagau}, \citenamefont {Bravaya}, \citenamefont {Epifanovsky},
  \citenamefont {Shao}, \citenamefont {Sundstrom}, \citenamefont
  {Head-Gordon},\ and\ \citenamefont {Krylov}}]{Zuev14}%
  \BibitemOpen
  \bibfield  {author} {\bibinfo {author} {\bibfnamefont {D.}~\bibnamefont
  {Zuev}}, \bibinfo {author} {\bibfnamefont {T.-C.}\ \bibnamefont {Jagau}},
  \bibinfo {author} {\bibfnamefont {K.~B.}\ \bibnamefont {Bravaya}}, \bibinfo
  {author} {\bibfnamefont {E.}~\bibnamefont {Epifanovsky}}, \bibinfo {author}
  {\bibfnamefont {Y.}~\bibnamefont {Shao}}, \bibinfo {author} {\bibfnamefont
  {E.}~\bibnamefont {Sundstrom}}, \bibinfo {author} {\bibfnamefont
  {M.}~\bibnamefont {Head-Gordon}}, \ and\ \bibinfo {author} {\bibfnamefont
  {A.I.}\ \bibnamefont {Krylov}},\ }\bibfield  {title} {\enquote {\bibinfo
  {title} {Complex absorbing potentials within {EOM-CC} family of methods:
  {T}heory, implementation, and benchmarks},}\ }\href {\doibase
  10.1063/1.4885056} {\bibfield  {journal} {\bibinfo  {journal} {J. Chem.
  Phys.}\ }\textbf {\bibinfo {volume} {141}},\ \bibinfo {pages} {024102}
  (\bibinfo {year} {2014})}\BibitemShut {NoStop}%
\bibitem [{\citenamefont {Cederbaum}, \citenamefont {Domcke},\ and\
  \citenamefont {Schirmer}(1980)}]{Cederbaum80}%
  \BibitemOpen
  \bibfield  {author} {\bibinfo {author} {\bibfnamefont {L.~S.}\ \bibnamefont
  {Cederbaum}}, \bibinfo {author} {\bibfnamefont {W.}~\bibnamefont {Domcke}}, \
  and\ \bibinfo {author} {\bibfnamefont {J.}~\bibnamefont {Schirmer}},\
  }\bibfield  {title} {\enquote {\bibinfo {title} {Many-body theory of core
  holes},}\ }\href {\doibase 10.1103/PhysRevA.22.206} {\bibfield  {journal}
  {\bibinfo  {journal} {Phys. Rev. A}\ }\textbf {\bibinfo {volume} {22}},\
  \bibinfo {pages} {206–222} (\bibinfo {year} {1980})}\BibitemShut {NoStop}%
\bibitem [{\citenamefont {Coriani}\ and\ \citenamefont
  {Koch}(2015)}]{Coriani15}%
  \BibitemOpen
  \bibfield  {author} {\bibinfo {author} {\bibfnamefont {S.}~\bibnamefont
  {Coriani}}\ and\ \bibinfo {author} {\bibfnamefont {H.}~\bibnamefont {Koch}},\
  }\bibfield  {title} {\enquote {\bibinfo {title} {Communication: {X}-ray
  absorption spectra and core-ionization potentials within a core-valence
  separated coupled cluster framework},}\ }\href {\doibase 10.1063/1.4935712}
  {\bibfield  {journal} {\bibinfo  {journal} {J. Chem. Phys.}\ }\textbf
  {\bibinfo {volume} {143}},\ \bibinfo {pages} {181103} (\bibinfo {year}
  {2015})}\BibitemShut {NoStop}%
\bibitem [{\citenamefont {Vidal}\ \emph {et~al.}(2019)\citenamefont {Vidal},
  \citenamefont {Feng}, \citenamefont {Epi{Fano}vsky}, \citenamefont {Krylov},\
  and\ \citenamefont {Coriani}}]{Vidal19}%
  \BibitemOpen
  \bibfield  {author} {\bibinfo {author} {\bibfnamefont {M.~L.}\ \bibnamefont
  {Vidal}}, \bibinfo {author} {\bibfnamefont {X.}~\bibnamefont {Feng}},
  \bibinfo {author} {\bibfnamefont {E.}~\bibnamefont {Epifanovsky}}, \bibinfo
  {author} {\bibfnamefont {A.~I.}\ \bibnamefont {Krylov}}, \ and\ \bibinfo
  {author} {\bibfnamefont {S.}~\bibnamefont {Coriani}},\ }\bibfield  {title}
  {\enquote {\bibinfo {title} {New and efficient equation-of-motion
  coupled-cluster framework for core-excited and core-ionized states},}\ }\href
  {\doibase 10.1021/acs.jctc.9b00039} {\bibfield  {journal} {\bibinfo
  {journal} {J. Chem. Theory Comput.}\ }\textbf {\bibinfo {volume} {15}},\
  \bibinfo {pages} {3117–3133} (\bibinfo {year} {2019})}\BibitemShut
  {NoStop}%
\bibitem [{\citenamefont {Scheit}\ \emph {et~al.}(2004)\citenamefont {Scheit},
  \citenamefont {Averbukh}, \citenamefont {Meyer}, \citenamefont {Moiseyev},
  \citenamefont {Santra}, \citenamefont {Sommerfeld}, \citenamefont {Zobeley},\
  and\ \citenamefont {Cederbaum}}]{Scheit04}%
  \BibitemOpen
  \bibfield  {author} {\bibinfo {author} {\bibfnamefont {S.}~\bibnamefont
  {Scheit}}, \bibinfo {author} {\bibfnamefont {V.}~\bibnamefont {Averbukh}},
  \bibinfo {author} {\bibfnamefont {H.-D.}\ \bibnamefont {Meyer}}, \bibinfo
  {author} {\bibfnamefont {N.}~\bibnamefont {Moiseyev}}, \bibinfo {author}
  {\bibfnamefont {R.}~\bibnamefont {Santra}}, \bibinfo {author} {\bibfnamefont
  {T.}~\bibnamefont {Sommerfeld}}, \bibinfo {author} {\bibfnamefont
  {J.}~\bibnamefont {Zobeley}}, \ and\ \bibinfo {author} {\bibfnamefont
  {L.~S.}\ \bibnamefont {Cederbaum}},\ }\bibfield  {title} {\enquote {\bibinfo
  {title} {On the interatomic {Coulombic} decay in the {N}e dimer},}\ }\href
  {\doibase 10.1063/1.1794654} {\bibfield  {journal} {\bibinfo  {journal} {J.
  Chem. Phys.}\ }\textbf {\bibinfo {volume} {121}},\ \bibinfo {pages}
  {8393–8398} (\bibinfo {year} {2004})}\BibitemShut {NoStop}%
\bibitem [{\citenamefont {Jahnke}\ \emph {et~al.}(2004)\citenamefont {Jahnke},
  \citenamefont {Czasch}, \citenamefont {Schöffler}, \citenamefont
  {Schössler}, \citenamefont {Knapp}, \citenamefont {Käsz}, \citenamefont
  {Titze}, \citenamefont {Wimmer}, \citenamefont {Kreidi}, \citenamefont
  {Grisenti}, \citenamefont {Staudte}, \citenamefont {Jagutzki}, \citenamefont
  {Hergenhahn}, \citenamefont {Schmidt-Böcking},\ and\ \citenamefont
  {Dörner}}]{Jahnke04}%
  \BibitemOpen
  \bibfield  {author} {\bibinfo {author} {\bibfnamefont {T.}~\bibnamefont
  {Jahnke}}, \bibinfo {author} {\bibfnamefont {A.}~\bibnamefont {Czasch}},
  \bibinfo {author} {\bibfnamefont {M.~S.}\ \bibnamefont {Schöffler}},
  \bibinfo {author} {\bibfnamefont {S.}~\bibnamefont {Schössler}}, \bibinfo
  {author} {\bibfnamefont {A.}~\bibnamefont {Knapp}}, \bibinfo {author}
  {\bibfnamefont {M.}~\bibnamefont {Käsz}}, \bibinfo {author} {\bibfnamefont
  {J.}~\bibnamefont {Titze}}, \bibinfo {author} {\bibfnamefont
  {C.}~\bibnamefont {Wimmer}}, \bibinfo {author} {\bibfnamefont
  {K.}~\bibnamefont {Kreidi}}, \bibinfo {author} {\bibfnamefont {R.~E.}\
  \bibnamefont {Grisenti}}, \bibinfo {author} {\bibfnamefont {A.}~\bibnamefont
  {Staudte}}, \bibinfo {author} {\bibfnamefont {O.}~\bibnamefont {Jagutzki}},
  \bibinfo {author} {\bibfnamefont {U.}~\bibnamefont {Hergenhahn}}, \bibinfo
  {author} {\bibfnamefont {H.}~\bibnamefont {Schmidt-Böcking}}, \ and\
  \bibinfo {author} {\bibfnamefont {R.}~\bibnamefont {Dörner}},\ }\bibfield
  {title} {\enquote {\bibinfo {title} {Experimental observation of interatomic
  {Coulombic} decay in neon dimers},}\ }\href {\doibase
  10.1103/PhysRevLett.93.163401} {\bibfield  {journal} {\bibinfo  {journal}
  {Phys. Rev. Lett.}\ }\textbf {\bibinfo {volume} {93}},\ \bibinfo {pages}
  {163401} (\bibinfo {year} {2004})}\BibitemShut {NoStop}%
\bibitem [{\citenamefont {Schnorr}\ \emph {et~al.}(2013)\citenamefont
  {Schnorr}, \citenamefont {Senftleben}, \citenamefont {Kurka}, \citenamefont
  {Rudenko}, \citenamefont {Foucar}, \citenamefont {Schmid}, \citenamefont
  {Broska}, \citenamefont {Pfeifer}, \citenamefont {Meyer}, \citenamefont
  {Anielski}, \citenamefont {Boll}, \citenamefont {Rolles}, \citenamefont
  {Kübel}, \citenamefont {Kling}, \citenamefont {Jiang}, \citenamefont
  {Mondal}, \citenamefont {Tachibana}, \citenamefont {Ueda}, \citenamefont
  {Marchenko}, \citenamefont {Simon}, \citenamefont {Brenner}, \citenamefont
  {Treusch}, \citenamefont {Scheit}, \citenamefont {Averbukh}, \citenamefont
  {Ullrich}, \citenamefont {Schröter},\ and\ \citenamefont
  {Moshammer}}]{Schnorr13}%
  \BibitemOpen
  \bibfield  {author} {\bibinfo {author} {\bibfnamefont {K.}~\bibnamefont
  {Schnorr}}, \bibinfo {author} {\bibfnamefont {A.}~\bibnamefont {Senftleben}},
  \bibinfo {author} {\bibfnamefont {M.}~\bibnamefont {Kurka}}, \bibinfo
  {author} {\bibfnamefont {A.}~\bibnamefont {Rudenko}}, \bibinfo {author}
  {\bibfnamefont {L.}~\bibnamefont {Foucar}}, \bibinfo {author} {\bibfnamefont
  {G.}~\bibnamefont {Schmid}}, \bibinfo {author} {\bibfnamefont
  {A.}~\bibnamefont {Broska}}, \bibinfo {author} {\bibfnamefont
  {T.}~\bibnamefont {Pfeifer}}, \bibinfo {author} {\bibfnamefont
  {K.}~\bibnamefont {Meyer}}, \bibinfo {author} {\bibfnamefont
  {D.}~\bibnamefont {Anielski}}, \bibinfo {author} {\bibfnamefont
  {R.}~\bibnamefont {Boll}}, \bibinfo {author} {\bibfnamefont {D.}~\bibnamefont
  {Rolles}}, \bibinfo {author} {\bibfnamefont {M.}~\bibnamefont {Kübel}},
  \bibinfo {author} {\bibfnamefont {M.~F.}\ \bibnamefont {Kling}}, \bibinfo
  {author} {\bibfnamefont {Y.~H.}\ \bibnamefont {Jiang}}, \bibinfo {author}
  {\bibfnamefont {S.}~\bibnamefont {Mondal}}, \bibinfo {author} {\bibfnamefont
  {T.}~\bibnamefont {Tachibana}}, \bibinfo {author} {\bibfnamefont
  {K.}~\bibnamefont {Ueda}}, \bibinfo {author} {\bibfnamefont {T.}~\bibnamefont
  {Marchenko}}, \bibinfo {author} {\bibfnamefont {M.}~\bibnamefont {Simon}},
  \bibinfo {author} {\bibfnamefont {G.}~\bibnamefont {Brenner}}, \bibinfo
  {author} {\bibfnamefont {R.}~\bibnamefont {Treusch}}, \bibinfo {author}
  {\bibfnamefont {S.}~\bibnamefont {Scheit}}, \bibinfo {author} {\bibfnamefont
  {V.}~\bibnamefont {Averbukh}}, \bibinfo {author} {\bibfnamefont
  {J.}~\bibnamefont {Ullrich}}, \bibinfo {author} {\bibfnamefont {C.~D.}\
  \bibnamefont {Schröter}}, \ and\ \bibinfo {author} {\bibfnamefont
  {R.}~\bibnamefont {Moshammer}},\ }\bibfield  {title} {\enquote {\bibinfo
  {title} {Time-resolved measurement of interatomic {Coulombic} decay in
  {Ne$_2$}},}\ }\href {\doibase 10.1103/PhysRevLett.111.093402} {\bibfield
  {journal} {\bibinfo  {journal} {Phys. Rev. Lett.}\ }\textbf {\bibinfo
  {volume} {111}},\ \bibinfo {pages} {093402} (\bibinfo {year}
  {2013})}\BibitemShut {NoStop}%
\bibitem [{\citenamefont {Scheit}\ \emph {et~al.}(2006)\citenamefont {Scheit},
  \citenamefont {Averbukh}, \citenamefont {Meyer}, \citenamefont {Zobeley},\
  and\ \citenamefont {Cederbaum}}]{Scheit06}%
  \BibitemOpen
  \bibfield  {author} {\bibinfo {author} {\bibfnamefont {S.}~\bibnamefont
  {Scheit}}, \bibinfo {author} {\bibfnamefont {V.}~\bibnamefont {Averbukh}},
  \bibinfo {author} {\bibfnamefont {H.-D.}\ \bibnamefont {Meyer}}, \bibinfo
  {author} {\bibfnamefont {J.}~\bibnamefont {Zobeley}}, \ and\ \bibinfo
  {author} {\bibfnamefont {L.~S.}\ \bibnamefont {Cederbaum}},\ }\bibfield
  {title} {\enquote {\bibinfo {title} {Interatomic {Coulombic} decay in a
  heteroatomic rare gas cluster},}\ }\href {\doibase 10.1063/1.2185637}
  {\bibfield  {journal} {\bibinfo  {journal} {J. Chem. Phys.}\ }\textbf
  {\bibinfo {volume} {124}},\ \bibinfo {pages} {154305} (\bibinfo {year}
  {2006})}\BibitemShut {NoStop}%
\bibitem [{\citenamefont {Santra}, \citenamefont {Cederbaum},\ and\
  \citenamefont {Meyer}(1999)}]{Santra99}%
  \BibitemOpen
  \bibfield  {author} {\bibinfo {author} {\bibfnamefont {R.}~\bibnamefont
  {Santra}}, \bibinfo {author} {\bibfnamefont {L.~S.}\ \bibnamefont
  {Cederbaum}}, \ and\ \bibinfo {author} {\bibfnamefont {H.~D.}\ \bibnamefont
  {Meyer}},\ }\bibfield  {title} {\enquote {\bibinfo {title} {Electronic decay
  of molecular clusters: non-stationary states computed by standard quantum
  chemistry methods},}\ }\href {\doibase 10.1016/S0009-2614(99)00226-2}
  {\bibfield  {journal} {\bibinfo  {journal} {Chem. Phys. Lett.}\ }\textbf
  {\bibinfo {volume} {303}},\ \bibinfo {pages} {413–419} (\bibinfo {year}
  {1999})}\BibitemShut {NoStop}%
\bibitem [{\citenamefont {Zobeley}, \citenamefont {Cederbaum},\ and\
  \citenamefont {Tarantelli}(1998)}]{Zobeley98}%
  \BibitemOpen
  \bibfield  {author} {\bibinfo {author} {\bibfnamefont {J.}~\bibnamefont
  {Zobeley}}, \bibinfo {author} {\bibfnamefont {L.~S.}\ \bibnamefont
  {Cederbaum}}, \ and\ \bibinfo {author} {\bibfnamefont {F.}~\bibnamefont
  {Tarantelli}},\ }\bibfield  {title} {\enquote {\bibinfo {title} {Highly
  excited electronic states of molecular clusters and their decay},}\ }\href
  {\doibase 10.1063/1.476448} {\bibfield  {journal} {\bibinfo  {journal} {J.
  Chem. Phys.}\ }\textbf {\bibinfo {volume} {108}},\ \bibinfo {pages}
  {9737–9750} (\bibinfo {year} {1998})}\BibitemShut {NoStop}%
\bibitem [{\citenamefont {Ghosh}, \citenamefont {Pal},\ and\ \citenamefont
  {Vaval}(2014)}]{Ghosh14-2}%
  \BibitemOpen
  \bibfield  {author} {\bibinfo {author} {\bibfnamefont {A.}~\bibnamefont
  {Ghosh}}, \bibinfo {author} {\bibfnamefont {S.}~\bibnamefont {Pal}}, \ and\
  \bibinfo {author} {\bibfnamefont {N.}~\bibnamefont {Vaval}},\ }\bibfield
  {title} {\enquote {\bibinfo {title} {Interatomic {Coulombic} decay in
  {HF}$_n$ (n = 2–3) clusters using {CAP/EOM-CCSD} method},}\ }\href
  {\doibase 10.1080/00268976.2013.852263} {\bibfield  {journal} {\bibinfo
  {journal} {Mol. Phys.}\ }\textbf {\bibinfo {volume} {112}},\ \bibinfo {pages}
  {669–673} (\bibinfo {year} {2014})}\BibitemShut {NoStop}%
\end{thebibliography}%

\end{document}